\documentclass[rnote]{aa} 
\usepackage{natbib,twoopt}
\usepackage[breaklinks=true]{hyperref} 
\bibpunct{(}{)}{;}{a}{}{,}             
\makeatletter
  \newcommandtwoopt{\citeads}[3][][]{\href{http://adsabs.harvard.edu/abs/#3}%
    {\def\hyper@linkstart##1##2{}%
     \let\hyper@linkend\@empty\citealp[#1][#2]{#3}}}
  \newcommandtwoopt{\citepads}[3][][]{\href{http://adsabs.harvard.edu/abs/#3}%
    {\def\hyper@linkstart##1##2{}%
     \let\hyper@linkend\@empty\citep[#1][#2]{#3}}}
  \newcommandtwoopt{\citetads}[3][][]{\href{http://adsabs.harvard.edu/abs/#3}%
    {\def\hyper@linkstart##1##2{}%
     \let\hyper@linkend\@empty\citet[#1][#2]{#3}}}
  \newcommandtwoopt{\citeyearads}[3][][]%
    {\href{http://adsabs.harvard.edu/abs/#3}
    {\def\hyper@linkstart##1##2{}%
     \let\hyper@linkend\@empty\citeyear[#1][#2]{#3}}}
\makeatother

\usepackage{graphicx}
\usepackage{relsize}
\usepackage{appendix}
\usepackage{gensymb}
\usepackage{siunitx}
\usepackage{microtype}
\usepackage[dvipsnames]{xcolor}
\usepackage[para,online,flushleft]{threeparttable}
\makeatletter
\def\TPT@doparanotes{\par
   \prevdepth\z@ \TPT@hsize
   \TPTnoteSettings
   \parindent\z@ \pretolerance 8
   \linepenalty 200
   \renewcommand\item[1][]{\relax\ifhmode \begingroup
       \unskip
       \advance\hsize 10em 
       \penalty -45 \hskip\z@\@plus\hsize \penalty-19
       \hskip .15\hsize \penalty 9999 \hskip-.15\hsize
       \hskip .01\hsize\@plus-\hsize\@minus.01\hsize 
       \hskip 0em\@plus .3em
      \endgroup\fi
      \tnote{##1}\,\ignorespaces}%
   \let\TPToverlap\relax
   \def\endtablenotes{\par}%
}

\usepackage{txfonts}
\usepackage{amsmath}
\usepackage[]{hyperref}
\hypersetup{
      colorlinks = true,
      citecolor ={black}
   }
\defcitealias{2020Dsilva}{Paper 1}
\defcitealias{2001vanderHucht}{VDH01}
\def\kms{km\,s$^{-1}$}
\def\fobs{$f_{\textrm{obs}}^{\textrm{WNE}}$}
\def\fobsWC{$f_{\textrm{obs}}^{\textrm{WC}}$}
\def\fint{$f_{\textrm{int}}^{\textrm{WNE}}$}
\def\fintWC{$f_{\textrm{int}}^{\textrm{WC}}$}

\def\DelRV{$\Delta$RV}

\def\Msun{$M_{\odot}$}

\def\Pmin{$P_\mathrm{min}$}

\def\logPmin{$\log P_\mathrm{min}$}
\def\logPmax{$\log P_\mathrm{max}$}

\newcommand{\NVred}{N\,{\sc v}\,$\lambda 4945$}
\newcommand{\NVblue}{N\,{\sc v} $\lambda \lambda 4604, 4620$}

\newcommand{\HeII}{He\,{\sc ii}\,$\lambda 4686$}

\newcommand{\NIVred}{N\,{\sc iv} $\lambda \lambda \lambda 7103, 7109, 7123$}

\newcommand{\niv}{N\,{\sc iv}}
\newcommand{\nv}{N\,{\sc v}}
\newcommand{\heii}{He\,{\sc ii}}

\begin{document}

   \title{A spectroscopic multiplicity survey of Galactic Wolf-Rayet stars\thanks{Based on observations made with the Mercator Telescope, operated on the island of La Palma by the Flemish Community, at the Spanish Observatorio del Roque de los Muchachos of the Instituto de Astrofísica de Canarias.}\fnmsep\thanks{Based on observations obtained with the HERMES spectrograph, which is supported by the Research Foundation - Flanders (FWO), Belgium, the Research Council of KU Leuven, Belgium, the Fonds National de la Recherche Scientifique (F.R.S.-FNRS), Belgium, the Royal Observatory of Belgium, the Observatoire de Genève, Switzerland and the Thüringer Landessternwarte Tautenburg, Germany.}}

   \subtitle{II. The northern WNE sequence}

   \author{K. Dsilva
          \inst{1},
          T. Shenar\inst{1,2}, 
          H. Sana\inst{1}
          \and
          P. Marchant\inst{1}
          }

   \institute{$^1$Institute of Astronomy, KU Leuven,
              Celestijnenlaan 200D, 3001 Leuven, Belgium\\
              $^2$Anton Pannekoek Institute for Astronomy, University of Amsterdam, Postbus 94249, 1090 GE Amsterdam, The Netherlands\\
              \email{karan.dsilva@kuleuven.be}
             }

   \date{Received: November 23, 2021; accepted: April 4, 2022}

 
  \abstract
   {Most massive stars reside in multiple systems that will interact over the course of their lifetime. This has important consequences on their future evolution and their end-of-life products. Classical Wolf-Rayet (WR) stars represent the final end stages of stellar evolution at the upper-mass end. While their observed multiplicity fraction is reported to be ${\sim}0.4$ in the Galaxy, their intrinsic multiplicity properties and the distributions of their orbital parameters remain insufficiently constrained to provide a reliable anchor to compare to evolutionary predictions.}
   {As part of a homogeneous, magnitude-limited ($V\leq12$) spectroscopic survey of northern Galactic WR stars, this paper aims to establish the observed and intrinsic multiplicity properties of the early-type nitrogen-rich WR population (WNE), including estimates of the multiplicity fraction and the shape of their orbital period distribution. Additionally, we compare these with the properties of the carbon-rich WR population (WC) stars obtained in the first paper of this series.}
   {We obtained high-resolution spectroscopic time series of the complete magnitude-limited sample of 16 WNE stars observable with the 1.2\,m Mercator telescope at La Palma, typically providing a time base of ${\text{about }}$two to eight years. We measured relative radial velocities (RVs) using cross-correlation and used RV variations to flag binary candidates. Using an updated Monte Carlo method with a Bayesian framework, we calculated the three-dimensional likelihood for the intrinsic binary fraction (\fint{}), the maximum period (\logPmax{}), and the power-law index for the period distribution ($\pi$) for the WNE population with \Pmin{} fixed at 1\,d. We also used this updated method to re-derive multiplicity parameters for the Galactic WC population.
   }
   {Adopting a peak-to-peak RV variability threshold of 50\,\kms{} as a criterion, we classify 7 of the 16 targets as binaries. This results in an observed multiplicity fraction (\fobs{}) of 0.44\,$\pm$\,0.12. Assuming flat priors, we derive the best-fit multiplicity properties \fint{}\,$= 0.56\substack{+0.20 \\ -0.15}$, \logPmax{}\,$= 4.60\substack{+0.40 \\ -0.77}$ , and $\pi$\,$= -0.30\substack{+0.55 \\ -0.53}$ for the parent WNE population. We explored different mass-ratio distributions and note that they did not change our results significantly. For the Galactic WC population from \citetalias{2020Dsilva}, we re-derive \fintWC{}\,$= 0.96\substack{+0.04 \\ -0.22}$, \logPmin{}\,$= 0.75\substack{+0.26 \\ -0.60}$, \logPmax{}\,$= 4.00\substack{+0.42 \\ -0.34}$ , and $\pi$\,$= 1.90\substack{+1.26 \\ -1.25}$. }
   {The derived multiplicity parameters for the WNE population are quite similar to those derived for main-sequence O binaries but differ from those of the WC population. The significant shift in the WC period distribution towards longer periods is too large to be explained via expansion of the orbit due to stellar winds, and we discuss possible implications of our results. Analysis of the WNL population and further investigation of various evolutionary scenarios is required to connect the different evolutionary phases of stars at the upper-mass end. }

   \keywords{stars: binaries --
                stars: Wolf-Rayet --
                techniques: cross-correlation
               }

   \maketitle
%

\section{Introduction}

Stars with initial masses $M_i \gtrsim 8\,M_{\odot}$ will end their lives through core-collapse supernovae or through direct collapse \citep{2003Heger}, leaving behind compact objects such as neutron stars or black holes. Through their strong stellar winds and energetic end-of-life explosions, these stars dominate the transport of energy and momentum into the interstellar medium \citep[see][]{2005MacLow,2019CrowtherHIIFeedback}. 

At the upper mass end, stars with $M_i \gtrsim 20\,M_{\odot}$ may enter the classical Wolf-Rayet (WR) phase, where after stripping (most of) their hydrogen-rich envelope, they reveal deep layers that are strongly enriched by products of nucleosynthesis \citep[for a review, see][]{2007Crowther}. The spectra of WR stars are characterised by strong, broad emission lines that result from strong, optically thick, radiation-driven winds. WR stars are generally understood to be the evolved, core He-burning counterparts of massive O stars and are immediate progenitors of black holes. WR stars play an important role in constraining the evolution of massive stars \citep{1998Vanbeveren, 2003MeynetMaeder,2016Shenar,2019Hamann,2019Sander,2019Woosley}, as well as the properties of supernovae and compact objects \citep{2014DeMink,2016Marchant,2018Hainich,2020Langer}.

WR stars are divided into three main spectral classes: the nitrogen (WN), carbon (WC), and oxygen (WO) sequences, which reflect the nitrogen-, carbon-, or oxygen-rich chemical composition of their atmospheres. \citet{1976Conti} first proposed that over the course of their main-sequence evolution, O stars that are massive enough may lose sufficient mass via stellar winds to reveal, first, the products of H-burning, followed by those of He-burning \citep[see also][]{1965Rublev}. These products are nitrogen and carbon, respectively, and the evolutionary counterparts are thought to be WN and WC stars. Ever since, this picture of evolution has been called the ``Conti scenario''. In reality, there is a fluid transition between O and WN stars \citep[Of, Of/WN, e.g.][]{1995ContiOf,1997CrowtherBohannanOf}, as well as WN and WC stars \citep[WN/WC stars,][]{1989ContiMassey}.


Wolf-Rayet stars can also be produced in close binary systems where the envelope of the star is stripped through Roche-lobe overflow by a companion, or through a combination of stellar winds and mass transfer \citep{1967Paczynski}. The dominant formation channel for WR stars as a function of mass and metallicity is still widely debated \citep{1998Vanbeveren,2014Neugent,2019Shenar,2020Shenar}. As the immediate progenitors of black holes and neutron stars, the characteristics of WR stars directly determine the observed properties of supernovae \citep{2017Zapartas,2018EldridgeCurvePop} as well as the expected distribution of gravitational wave systems \citep{2019Eldridge}.

Radial velocity (RV) surveys in the last few decades have shown that the majority of main-sequence O stars are in multiple systems with periods of up to a few years, with a significant fraction at periods shorter than a few tens of days \citep{2009Mason,2012Sana,2013Sana,2014Sana,2014Kobulnicky,2014Barba,2016MaizApellaniz,2017Almeida,2019MaizApellaniz}. As they are likely to interact during their lifetime, comparing the multiplicity properties of O stars to that of WR stars during the WN and WC phases is crucial for calibrating the physics of binary interaction. Moreover, orbital analyses of WR binary systems allow us to measure black hole progenitor masses in a model-independent way.

The Galactic Catalogue of WR stars\footnote{http://pacrowther.staff.shef.ac.uk/WRcat/} (henceforth GCWR) is maintained by Paul Crowther \citep[originally, Appendix 1 in][]{2015RossloweCrowther}, with 667 objects currently (v1.25). Until recently, the observed multiplicity fraction of WR stars was reported to be ${\sim}0.4$ in the literature \citep[][henceforth VDH01]{2001vanderHucht}. However, there are multiple caveats to this value. Firstly, this fraction includes photometric, spectroscopic (with and without derived orbits), and visual WR binaries. Secondly, the number of WRs with known orbits remains small with respect to the number of WRs listed as binaries in the GCWR. Thirdly, \citetalias{2001vanderHucht} is a compilation of decades of work and involves different techniques, each with their own sensitivity and limitations. This means that correcting for observational biases is non-trivial. As a consequence, the intrinsic multiplicity fraction of WR stars and the distributions of orbital elements and mass ratios are insufficiently constrained. These are important resources that can be used to calibrate the physics in binary stellar evolution codes (e.g. efficiency of mass transfer and angular momentum transport) as well as recipes used in stellar population synthesis \citep[e.g. BPASS,][]{2017Eldridge}. 

In order to obtain a reliable correction for observational biases, accurate RV measurements, and their uncertainties, are of critical importance. With this in mind, a modern RV monitoring campaign was initiated in 2017 using the High-Efficiency and high-Resolution Mercator Echelle Spectrograph \citep[HERMES,][]{2011Raskin} mounted on the 1.2\,m Flemish Mercator Telescope on the island of La Palma. As described in \citet[][henceforth \citetalias{2020Dsilva}]{2020Dsilva}, the survey targets all WR stars brighter than $V=12$ that are visible from La Palma. In \citetalias{2020Dsilva}, we focused on the multiplicity properties of the WC population. Based on the 12 WC stars in our sample, we determined an observed binary fraction (\fobsWC{}) of $0.58\,\pm\,0.14,$ where uncertainties due to the sample size have been taken into account. Using Monte Carlo (MC) simulations and the measured RV uncertainties, we deduced the intrinsic multiplicity fraction (\fintWC{}) of the  Galactic WC population to be at least 0.72 at the 10\% significance limit. Moreover, most of these systems had \DelRV{} values that are indicative of orbital periods longer than a few hundred days, which is consistent with the small fraction (${\sim}10$\,\%) of known short-period WC binaries in \citetalias{2001vanderHucht}.

As a second paper in this series, we focus on 16 northern Galactic early-type WN (WNE) stars. Because WNE stars are believed to be the progenitors of WC stars, a direct connection between their multiplicity properties is expected. The sample, observations, and data reduction are presented in Sect. \ref{sect:sample}. Section \ref{sect:RVdet} elaborates on the RV measurements and relevant masks used for cross-correlation. In Sect. \ref{sect:results} we discuss the spectroscopic binary fraction and compare it to the literature. In Sect. \ref{sect:intbinfrac} we discuss the correction for observational biases and the intrinsic binary fraction. In Sect. \ref{sect:evolution} we discuss the evolutionary consequences. Section \ref{sect:conclusions} presents our conclusions.

\section{Sample and data reduction} \label{sect:sample}
\subsection{Sample}

\begin{figure}
    \centering
    \includegraphics[width=0.45\textwidth]{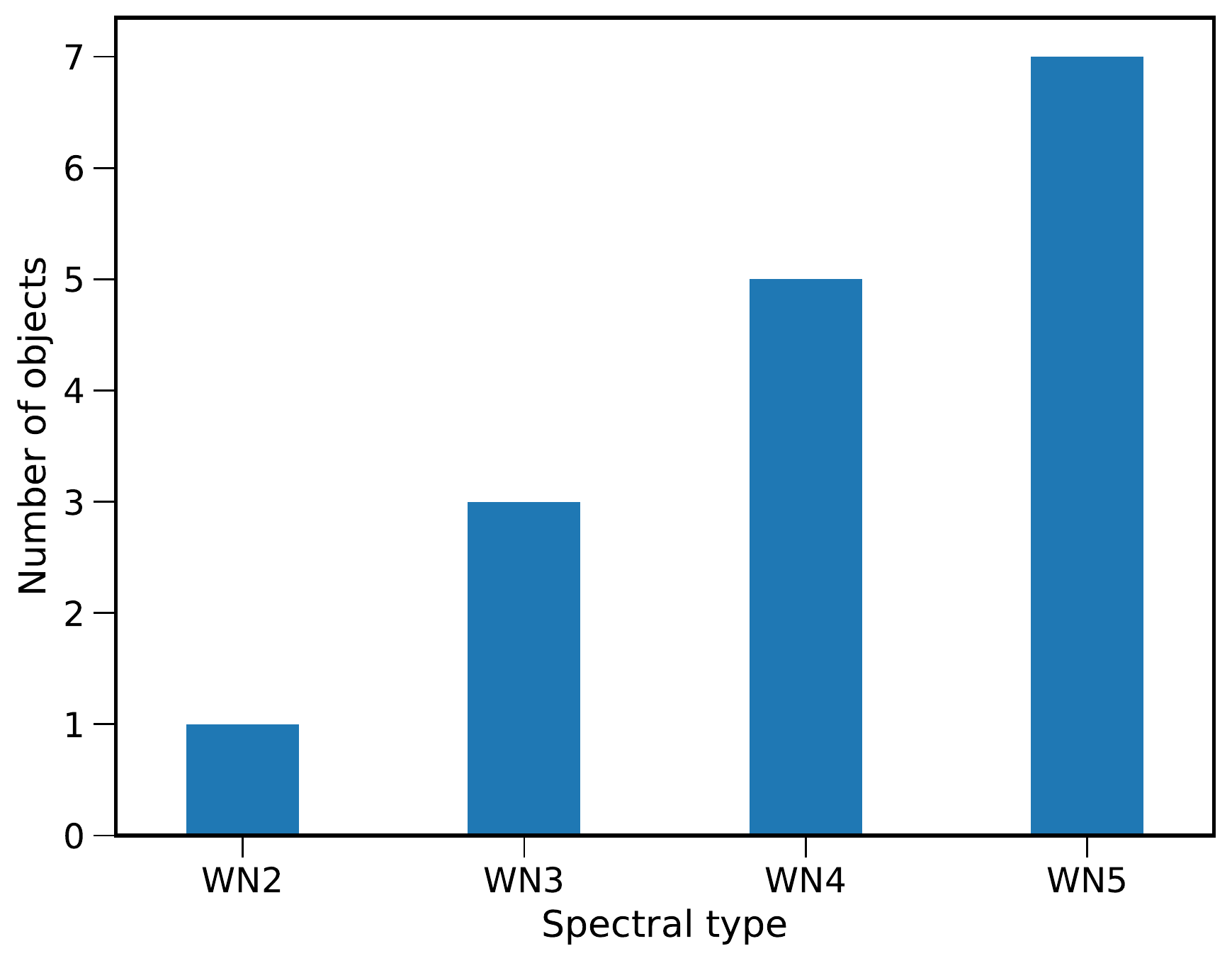}
    \caption{Distribution of spectral types within the present WNE sample.}
    \label{fig:target_dist}
\end{figure}

This work focuses on the WNE stars in our sample, which we define as WN stars with spectral types equal to or earlier than WN5. In cases where objects are classified as `WN5-WN6' in the GCWR, we consider them as WN5 for simplicity. From the GCWR we selected the abovementioned stars with $V\,\leq\,12$ and declination $\delta \ge$ $-30\degree$. For objects without a $V$ -band magnitude, we used narrow-band $\varv$ magnitudes \citepads{1968bSmith,1984Massey}, with the condition $\varv\,\leq\,13$. This resulted in a magnitude-limited sample of 16 WNE stars whose known properties are summarised in Table \ref{tab:WN_data}. Their spectral-type distribution is shown in Fig. \ref{fig:target_dist}.

Observations were carried out with the HERMES spectrograph mounted on the 1.2\,m Mercator telescope at the Roque de los Muchachos observatory in La Palma, Spain. HERMES covers the wavelength range from 3800\,\r{A} to 9000\,\r{A} with a spectral resolving power of R\,$=\lambda/\Delta\lambda\,{\sim}$\,85000. We obtained at least six epochs of each of the target stars. When available, we used archival HERMES observations in the RV analysis, which reach a typical time base of two to eight years. The number of spectra and the sampling time of the observations are shown in Table \ref{tab:wr_epochs}.  

\subsection{Data reduction and normalisation}
The data reduction procedures for our spectra are comprehensively described in \citetalias{2020Dsilva}. After the standard reduction from the HERMES pipeline \citep{2011Raskin}, a correction for the instrumental response function is applied. Molecfit \citep{2015SmetteMolecfit,2015KauschMolecfit} is then used to correct the spectra for telluric contamination. After this step, what remains is the effect of interstellar reddening superimposed on the continuum from the wind. The spectra are then normalised in a homogeneous way that minimises its effect on the RV measurements.

We used a model of a normalised spectrum of a WNE star from the Potsdam Wolf-Rayet code \citep[PoWR:][]{2002Grafener,2003HamannGrafenerPoWR,2004HamannGrafenerWN,2015TodtWNmodels} to identify pseudo-continuum regions, in the blue around 5300\,\r{A} and in the red around 8100\,\r{A}. We then anchored the stellar continuum model for the same WNE star to the red point and applied a reddening to fit the slope. We explored various reddening laws and R$_V$ values and realised that they did not significantly impact the homogeneity of the normalisation process. Therefore, we used the reddening law from \citetads{2004Fitzpatrick} and R$_V$ of 3.1, the average for the galaxy. 

\begin{table}
\centering
\caption{List of WNE stars in our RV monitoring campaign, providing the number of spectra, time coverage, and average signal-to-noise (S/N) per resolution element at \SI{4400}{\angstrom} ($\Delta\lambda\,\simeq\,$0.05 \r{A}).}
\begin{tabular}{cccc}
\hline \hline
WR\#&Spectra&Time coverage (d)&Average S/N \\ \hline
1&27&1263.8&40 \\
2&18&1084.9&45 \\
3&8&119.8&70 \\
6&28&1025.2&120 \\
7&7&642.3&45 \\
10&6&678.1&60 \\
110&11&1141.9&50 \\
127&11&2645.8&80 \\
128&10&1119.9&80 \\
133&21&2890.0&200 \\
138&40&2898.1&100 \\
139&34&2953.9&150 \\
141&14&2140.1&50 \\
151&6&924.3&20 \\
152&7&921.3&40 \\
157&12&2168.1&70 \\ \hline

\end{tabular}
\label{tab:wr_epochs}
\end{table}
\section{Measuring radial velocities}\label{sect:RVdet}
Deriving RVs of WR stars is not a trivial process given their strong, broad emission lines that form in their outflowing winds. Of the methods that are used extensively in the community, we used cross-correlation as our tool of choice, with a statistical framework that allows us to derive meaningful uncertainties on our RV measurements \citep[][see \citetalias{2020Dsilva} for more details]{2003Zucker}. As mentioned in \citetalias{2020Dsilva}, we focus on the RVs of the WR star even in SB2 systems to maintain homogeneity in the analysis and to focus on binary detection instead of orbital analysis. We briefly describe the formulation and assumptions of the cross-correlation function (CCF). 

\subsection{Cross-correlation and mask selection}
The CCF is the convolution of a chosen template to the data to measure the velocity shift. It is represented by a log-likelihood function, where the maximum is the RV of the spectrum. The uncertainty on the RV measurement can then be derived from maximum log-likelihood theory. An implicit assumption of the method is that the template accurately reproduces the data. Because WR models often do not to accurately represent the shape of both strong and weak lines simultaneously, we chose the highest signal-to-noise (S/N) epoch as the template. The measured RVs were then used to construct a weighted co-added spectrum of a higher S/N. This process was iterated until the derived RVs no longer changed significantly within the measurement errors, at which point the final template and measurements were converged. The S/N of the template contributes significantly to the measured uncertainty, and this was minimised by creating a high S/N co-added spectrum. A caveat is that the RVs derived in this fashion are relative RVs. While this is sufficient for binary detection, an absolute shift must be applied when comparing to RVs from the literature. 

When compared with the WC stars analysed in \citetalias{2020Dsilva}, the WNE stars in our sample exhibit stronger line-profile variability in their spectra. Therefore, the assumption that the template accurately represents the data often breaks down for variable stars. This directly results in larger uncertainties on the derived RVs. In order to ensure that the templates can represent the data as accurately as possible, for each star we selected encompassing spectral lines that are least affected by line-profile variability. Depending on how variable the stars are, we used lines of \niv, \nv, \heii,{} and when possible, Balmer lines. In cases of strong line-profile variability, we tend to only use \nv{} lines and occasionally \niv{} lines because they are thought to form in the hotter, inner regions of the stellar wind. We used a combination of all the different lines of various strengths and ions used to measure RVs and call this mask `full spec' (e.g. a combination of \nv{}, \niv,{} and \heii{} lines).

\begin{figure}
    \centering
    \includegraphics[width=0.49\textwidth]{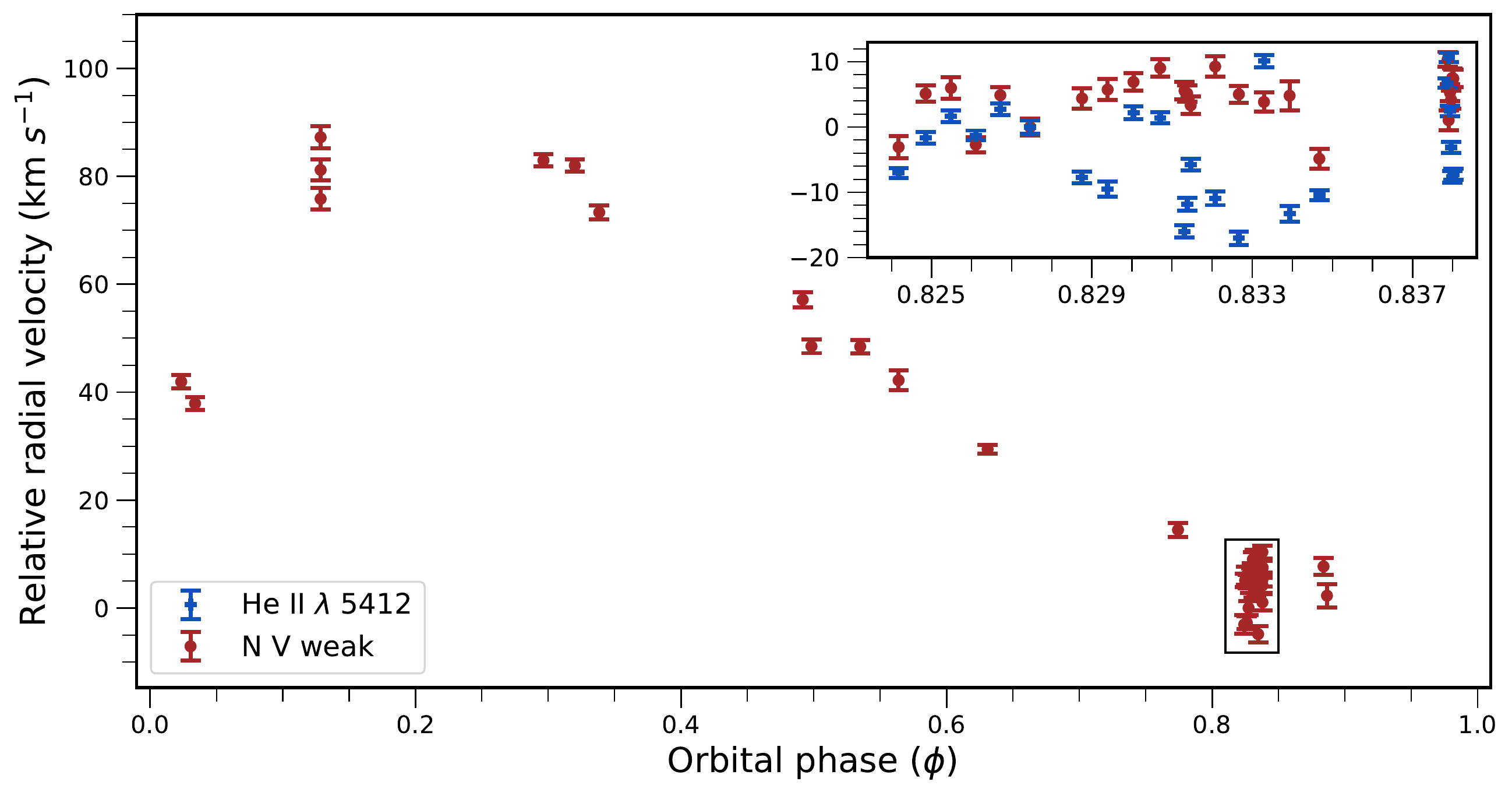}
    \caption{Phased RV plot for WR 138 with HERMES data. The inset shows the high-cadence measurements (black box) with the \nv{} weak lines and the He\,{\sc ii}\,$\lambda 5412$ line. The ephemeris used for the periastron passage was JD 2445284 \citep{1990Annuk}, with the period of 1521.2\,$\pm$\,35\,d from \citet{2013Palate}.}
    \label{fig:WR138_phased}
\end{figure}

\begin{figure}[t]
    \centering
    \includegraphics[width=0.49\textwidth]{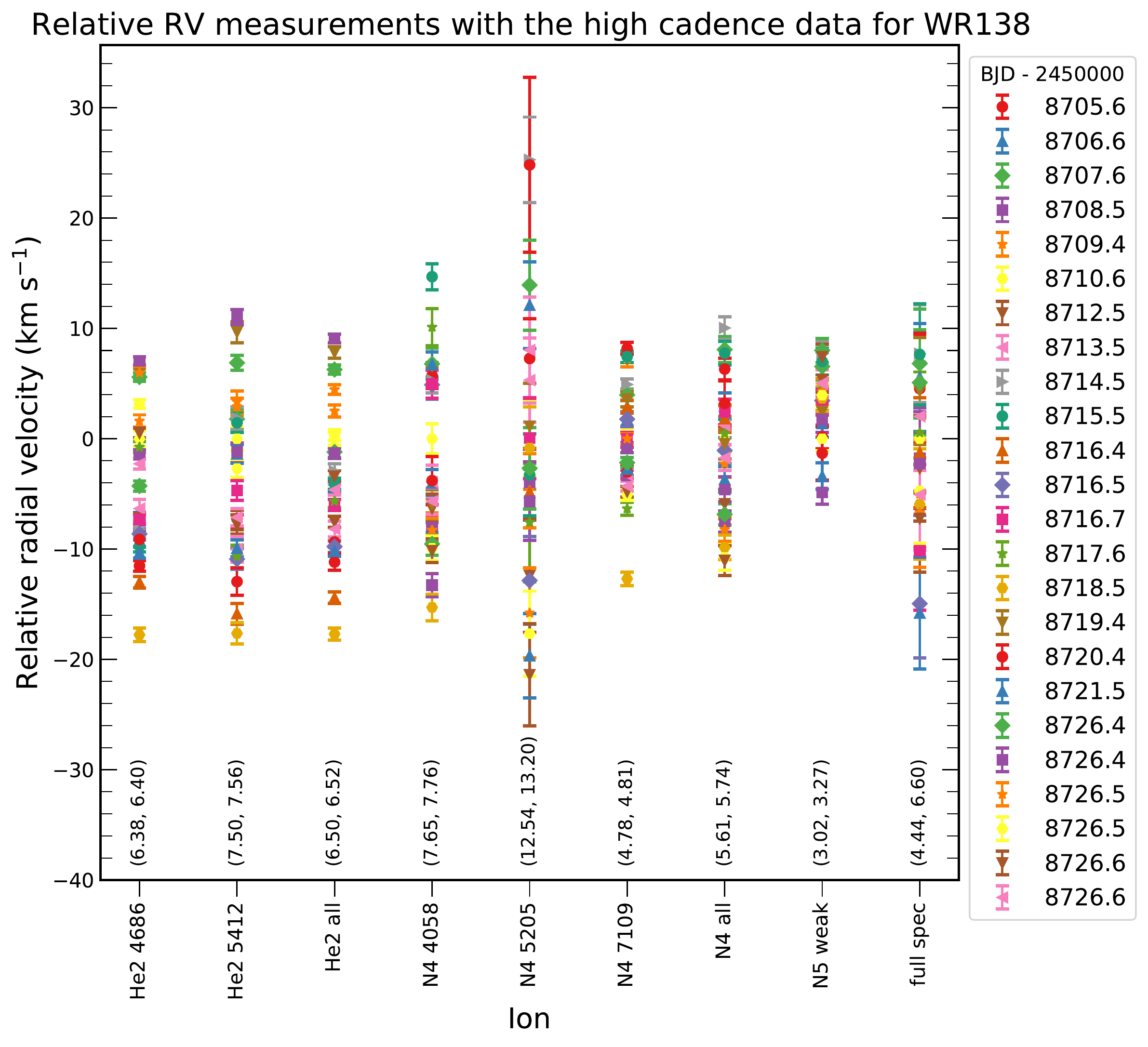}
    \caption{High-cadence relative RV measurements for WR 138 using different masks. At the bottom of the plot are the values of ($\sigma_w$, $\sigma_{\textrm{RV}}$) (\kms{}) for each mask. As mentioned in the text, `full spec' is a combination of all the masks to its left and not the entire spectrum. The \niv{} line at 5205\,\r{A} is very weak, which explains its high scatter.}
    \label{fig:sc_ions}
\end{figure}
\subsection{Studying the long-period binary system WR 138 to quantify wind variability} \label{sect:windVariability}
 
In \citetalias{2020Dsilva}, we analysed high-cadence data of the long-period binary system WR 137 in order to understand the effect of wind variability and other systematic effects on the RV measurements. As a nitrogen-rich analogue, we studied the long-period binary WR 138 (WN5o\,+\,O9V) with a period of 4.16\,yr and a low eccentricity of 0.3 \citep{2013Palate,1990Annuk,2016Richardson}. \citet{2013Palate} also studied the system in X-rays, concluding that its emission is normal for a colliding-wind WR\,+\,OB binary system, although the sampling of their data did not allow them to check for the existence of phase-locked variability of the X-ray emission. 
 
With this in mind, we collected high-cadence HERMES data in August of 2019 (with the periastron passage in June 2020). In total, we obtained 18 spectra over 16 consecutive nights and 6 spectra over one night a few days later (Table \ref{tab:WR138}). The phased RV plot of WR 138 is shown in Fig. \ref{fig:WR138_phased} with an emphasis on the high-cadence data over different masks. Because the high-cadence data were collected at the orbital phase of 0.8, the effect of wind-wind collision \citep[e.g.][]{2002Luehrs} on the measured RVs could be significant. Using the systemic and stellar parameters \citep{1990Annuk,2016Richardson}, we computed the cooling parameter $\chi$ \citep{1992Stevens} and found the shocked region to be adiabatic. As the cooling timescale is longer than the escape timescale, the contribution to recombination lines in the optical is minimal. Therefore, the contribution of wind-wind collision to the RV measurements should not change significantly over the timescale of the high-cadence observations. Even if wind-wind collision were to contribute significantly to the RV measurements, the derived scatter in the high-cadence study would be an upper limit on the intrinsic wind variability of the WR star.

    
For the high-cadence data, the contributions to the uncertainties in the measured RVs are statistical (i.e. from the method, $\sigma_p$) and physical (i.e. from the wind physics, $\sigma_w$). The weighted standard deviation of the high-cadence measurements are indicative of the true error, which we denote by $\sigma_{\textrm{RV}}$ (eq. 12 in \citetalias{2020Dsilva}). Assuming $\sigma_p$ and $\sigma_w$ are uncorrelated, they can simply be added quadratically as

\begin{equation} \label{eq:sigmaRV}
    \sigma_{\textrm{RV}} = \sqrt{\sigma_w^2 + \sigma_p^2}.
\end{equation}

For each mask, $\sigma_p$ is calculated from the co-variance matrix of the CCF (eq. 9 in \citetalias{2020Dsilva}). Using Eq. \ref{eq:sigmaRV} along with the calculated values of $\sigma_p$ and $\sigma_{\textrm{RV}}$, we obtain an estimate for $\sigma_w$. The high-cadence data therefore allow us to isolate and quantify the effect of wind variability on the RV measurements. The lines that are least affected by wind variability are \nv{} weak lines (App. \ref{apdx:comments}, $\sigma_w{\sim}\,5$\,\kms{}). Lines of \heii{} and \niv{} lines are found to be more variable ($\sigma_w{\sim}\,10$\,\kms{}). As their formation regions can be spatially different, these measurements can be helpful probes of turbulence and other physical processes. The difference in dispersion for lines of \niv{} arise from the different line strengths and since the S/N in the blue is much lower than in the red. The mask `full spec' is a combination of the \niv{}, \heii,{} and \nv{} lines. The mask `N4 7109' covers the triplet \NIVred{}. Ultimately, we chose to measure RVs for WR 138 using weak \nv{} lines, namely a combination of \NVblue,{} and \NVred{}.

Although we observed that weak \nv{} lines showed the least intrinsic variability for WR 138, this  cannot necessarily be generalised for all WNE stars. However, for all the objects in our sample, we realised that the \nv{} lines indeed had the least RV dispersion, and we therefore used them to measure RVs. For stars with strong line-profile variability (WR 2, WR 6, WR 7, and WR 110), we used the combination of all masks (`full spec') to measure RVs (App.\,\ref{apdx:comments}). The full journal of RV measurements for each object with their respective masks is given in App.\,\ref{s:tables_RV}.

\begin{figure}[t]
    \centering
    \includegraphics[width=0.49\textwidth]{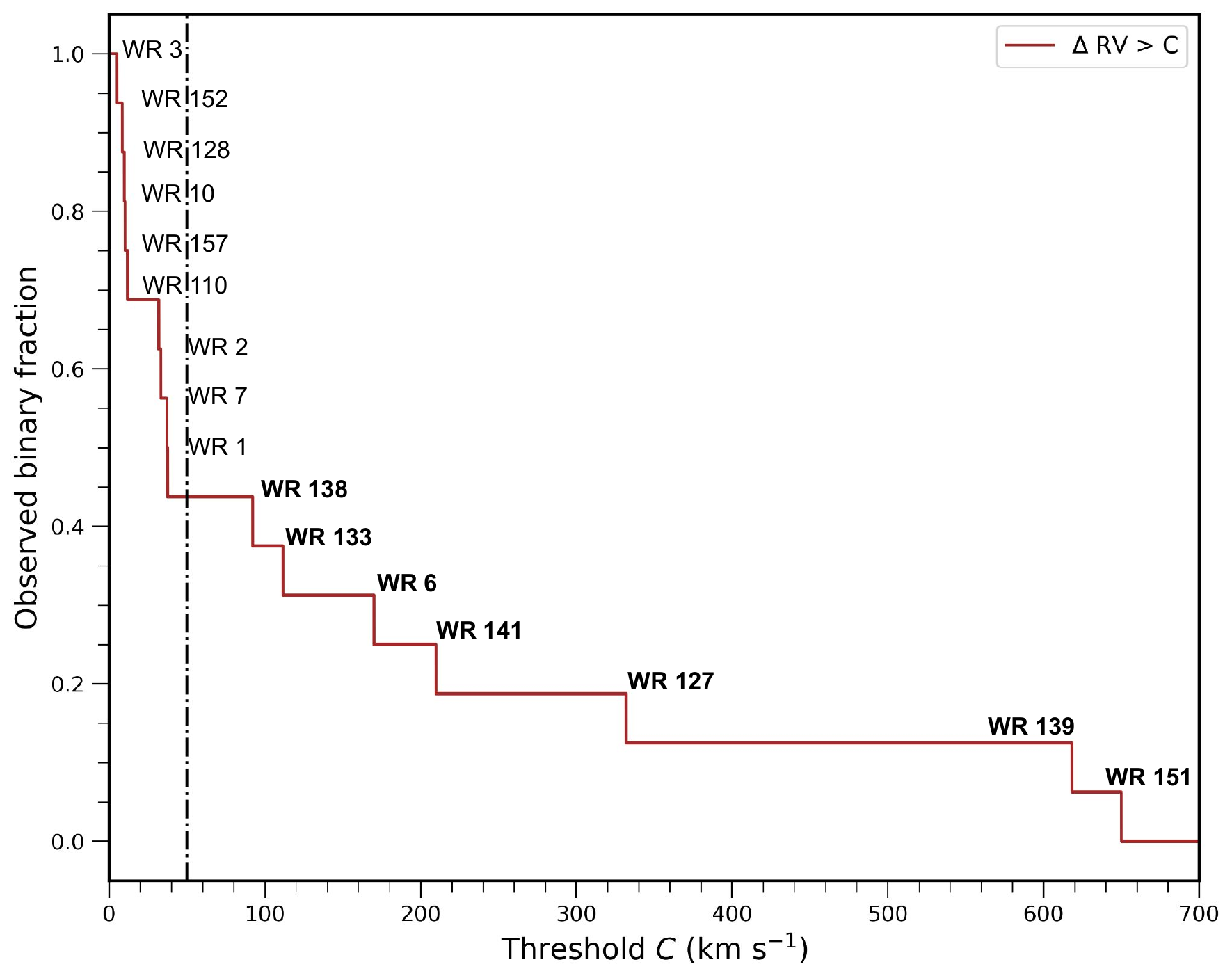}
    \caption{Non-parametric threshold plot showing the inverse cumulative distribution of peak-to-peak RV variability in our sample, which is equivalent to the observed binary fraction (\fobs{}) as a function of the adopted peak-to-peak RV variability threshold $C$. Names in bold are known binaries with established spectroscopic periods for the WR components.}
    \label{fig:binfrac}
\end{figure}

\begin{table*}
\centering
\caption{Overview of the known multiplicity properties of our sample of WNE stars together with the results of our RV measurements. $\Delta$ RV and $\sigma_{\textrm{RV}}$ are calculated in this work and are used to identify tentative spectroscopic binaries. The spectral types are taken from the GCWR unless indicated otherwise. The binary status of this work is reported based on the spectroscopic observations. }

\begin{threeparttable}
\begin{tabular}{ccccccccc}
\hline \hline

WR\# & Spectral Type & \multicolumn{2}{c}{Binary Status} & Period & e & $\Delta$ RV  & $\sigma_{\textrm{RV}}$ & \DelRV{} $>$ C\\ 
 & (GCWR) & (GCWR) & This work & (d) & &(\SI{}{\km\per\s}) &(\SI{}{\km\per\s}) &  \\\hline 
1 & WN4 & SB1? & - & - & - & 37.5 & 8.6 & no\\
2 & WN2 & VB & VB\tnote{(l)} & - & - & 36.9 & 8.3 & no\\
3 & WN3ha & SB2 & - & -\tnote{(a)} & - & 5.0 & 1.8 & no\\
6 & WN4b & SB1 & SB1 & 3.8\tnote{(b)} & 0.1\tnote{(b)} & 170.0 & 40.3 & yes\\
7 & WN4b & - & - & - & - & 33.3 & 10.8 & no\\
10 & WN5h & VB & VB\tnote{(l)} & - & - & 9.9 & 3.6 & no\\
110 & WN5b & - & - & - & - & 31.8 & 11.3 & no\\
127 & WN3b\,+\,O9.5V & SB2 & SB2 & 9.5550\,$\pm$\,0.0002\tnote{(c)} & 0.04\,$\pm$\,0.02\tnote{(c)} & 332.0 & 108.9 & yes\\
128 & WN4(h) & SB2?, VB & VB\tnote{(l)} & - & - & 9.7 & 2.8 & no\\
133 & WN5o\,+\,O9I & SB2, VB & SB2, VB\tnote{(l)} & 112.780\,$\pm$\,0.036\tnote{(d)} & 0.3558\,$\pm$\,0.0050\tnote{(d)} & 111.6 & 25.8 & yes\\
138 & WN5o\,+\,O9V\tnote{(e)} & SB2, VB & SB2, VB\tnote{(l)} & 1521.2\,$\pm$\,35\tnote{(f)} & 0.3\tnote{(g)} & 92.1 & 29.4 & yes\\
139 & WN5o\,+\,O6III-V & SB2 & SB2 & 4.212435\tnote{(h)} & 0.036\,$\pm$\,0.009\tnote{(h)} & 618.3 & 168.9 & yes\\
141 & WN5o\,+\,O5V-III & SB2 & SB2 & 21.6895\,$\pm$\,0.0003\tnote{(i)} & 0.018\,$\pm$\,0.035\tnote{(i)} & 209.8 & 66.5 & yes\\
151 & WN4o\,+\,O5V & SB2 & SB2 & 2.12691\,d\tnote{(j)} & 0 (fixed)\tnote{(j)} & 650.0 & 220.7 & yes\\
152 & WN3(h) & - & - & - & - & 8.3 & 2.6 & no\\
157 & WN5o\,+\,(OB\,+\,OB)\tnote{(k)} & VB & SB3, VB\tnote{(l)} & - & - & 11.8 & 3.6 & no\\
\hline 
\end{tabular}
\begin{tablenotes}[para]
    \item[(a)] \citet{1986Moffat+abs} reported a period of 46.8\,d, but it is not supported by our analysis (App. \ref{apdx:comments}), \\
    \item[(b)] \citet{2020Koenigsberger},
    \item[(c)] \citet{2011delachevrotiere} ,
    \item[(d)] \citet{2021Richardson},
    \item[(e)] \citet{2016Richardson},\\
    \item[(f)] \citet{2013Palate},
    \item[(g)] \citet{1990Annuk},
    \tnote{(h)} \citet{1994Marchenko}\,[fixed P; 4.212435\,d],
    \tnote{(i)} \citet{1998Marchenko1998WR141},\\
    \tnote{(j)} \citet{2009HuttonWR151}, 
    \tnote{(k)} This work, 
    \tnote{(l)} VB status carried over from the GCWR.

\end{tablenotes}
\end{threeparttable}
\label{tab:WN_data}
\end{table*}

\section{Observed binary fraction}\label{sect:results}

For each star in our sample, we measured the RVs and obtained their peak-to-peak variability amplitude (\DelRV{}). We then selected a threshold to determine the boundary beyond which the RV amplitude would be dominated by orbital motion in a binary system. We plot the binary fraction as a function of this threshold in a non-parametric threshold plot (Fig.\,\ref{fig:binfrac}). By moving the threshold from higher to lower RV values (right to left in Fig.\,\ref{fig:binfrac}), we classify more stars as RV variable, hence putative spectroscopic binaries. A summary of the characteristics of the WNE stars in our sample can be found in Table \ref{tab:WN_data}. As we cannot verify the status of objects classified as visual binaries (VB) in the GWCR, we simply carry over their status in our classification.

The next step was to choose an adequate value for the threshold $C$ such that stars with \DelRV{}$\ge$\,$C$ were classified as binaries. It is important that the chosen threshold avoids false positives, in particular, those that are erroneously classified as binaries due to strong intrinsic variability. The high-cadence study of WR 138 is a useful guideline to ensure this. The peak-to-peak RV amplitude measured due to wind variability was observed to be 15\,\kms{} for the \nv{} lines (Fig. \ref{fig:sc_ions}). We therefore chose a conservative threshold larger than three times this value, that is, 50\,\kms{}. This results in 7 of the 16 observed binaries in the sample, \fobs{} $= 0.44$ with a binomial error of $\pm$\,0.12. Interestingly, all the known spectroscopic binaries in our sample with an established orbital solution are properly identified with the adopted threshold. For objects with \DelRV{}\,$<$\,50\,\kms{} and a large enough RV time series (WR 1 and WR 2), we investigated possible periodicities using a Lomb-Scargle periodogram \citep{1976Lomb,1982Scargle}, but found no coherent periodicity.

\section{Intrinsic multiplicity properties}\label{sect:intbinfrac}
\subsection{Detection probability} \label{sect:detection_probability}
In order to determine the intrinsic multiplicity fraction of Galactic WNE stars, \fint{}, it is important to understand the observational biases of the RV campaign. The number of detected binaries is merely a minimum estimate of the true number of  WNE binaries in our sample. Physical and orbital properties of the systems, geometrical effects, time coverage, and sampling of the RV study all affect the binary detection probability. As in \citetalias{2020Dsilva}, we modelled the detection probability, estimated the number of undetected binaries, and constrained some of the properties of the parent WNE population.

Using MC simulations, we determined the sensitivity of our survey with the method explained in \citet{2019Patrick}. We constructed a grid with different orbital periods ($P$) and secondary masses ($M_2$) between 1 and 10$^5$\,d and 1 and 40\,\Msun{} , respectively. The grid was evenly spaced with steps of $\Delta \log (P/\textrm{d}) = 0.1$ and $\Delta M_2 = 1$\,\Msun{}. For each grid point of $\log P$ and $M_2$, we simulated RV time series for 40\,000 binary systems assuming a primary mass of 20\,\Msun{}, which is similar to the typical masses around 10-30\,\Msun{} reported for WNs in the Galaxy \citep{2019Hamann}. The eccentricities were drawn from a flat distribution between 0.0 and 0.9. We assumed random orientations of the orbital plane in three-dimensional space and random times of periastron passage. 

\begin{figure}[t]
    \centering
    \includegraphics[width=0.48\textwidth]{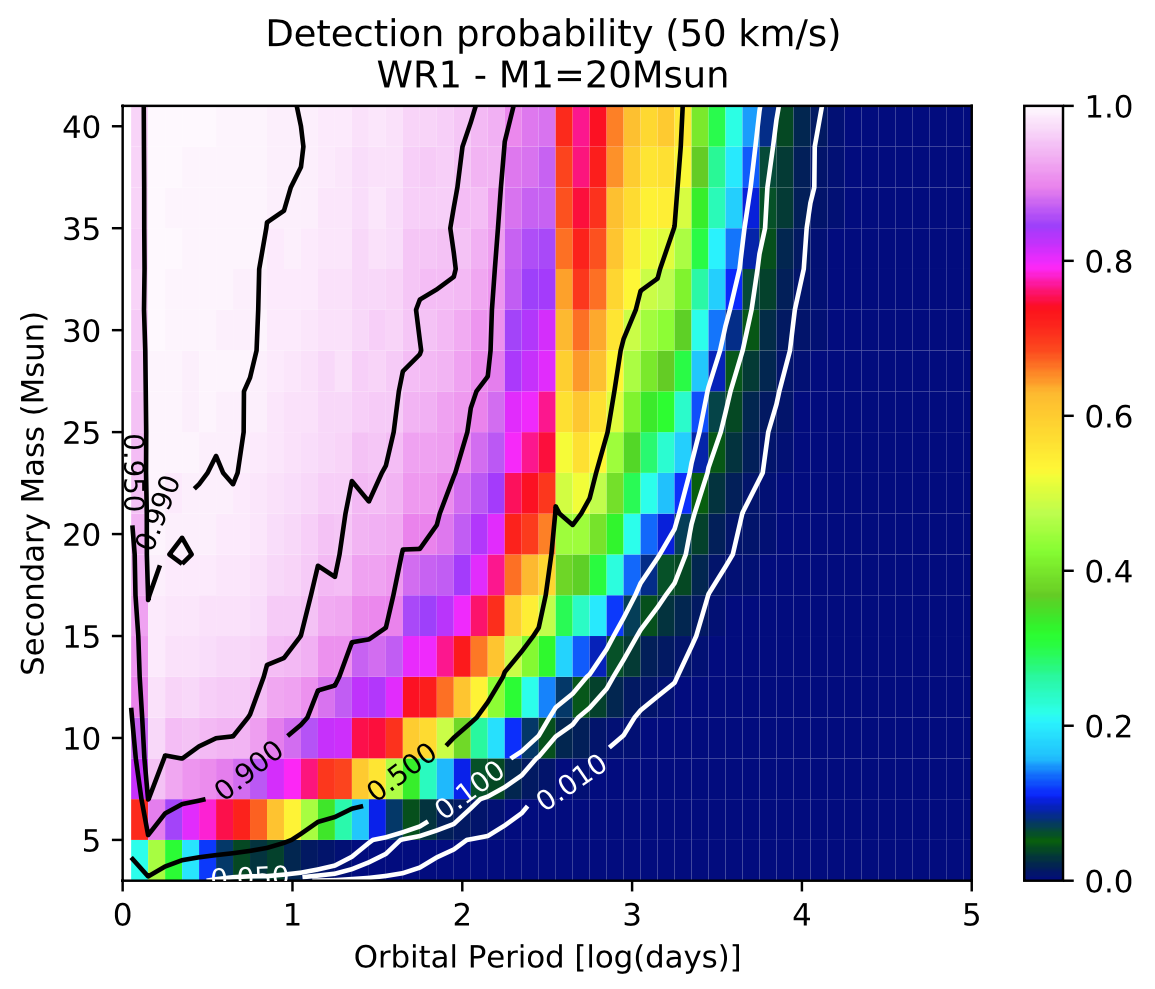}
    \caption{Binary detection probability for WR 1. It is computed assuming $M_1 = 20$\,\Msun{} and other conditions described in Sect. \ref{sect:detection_probability}. We have a detection probability of $\ge$90\% up to periods of 10$^2$\,d at $M_2\ge20$\,\Msun{} and $>{\sim}20$\% at a period of 10$^3$\,d.}
    \label{fig:Pdetect_WR1}
\end{figure}

\begin{figure}[t]
    \centering
    \includegraphics[width=0.48\textwidth]{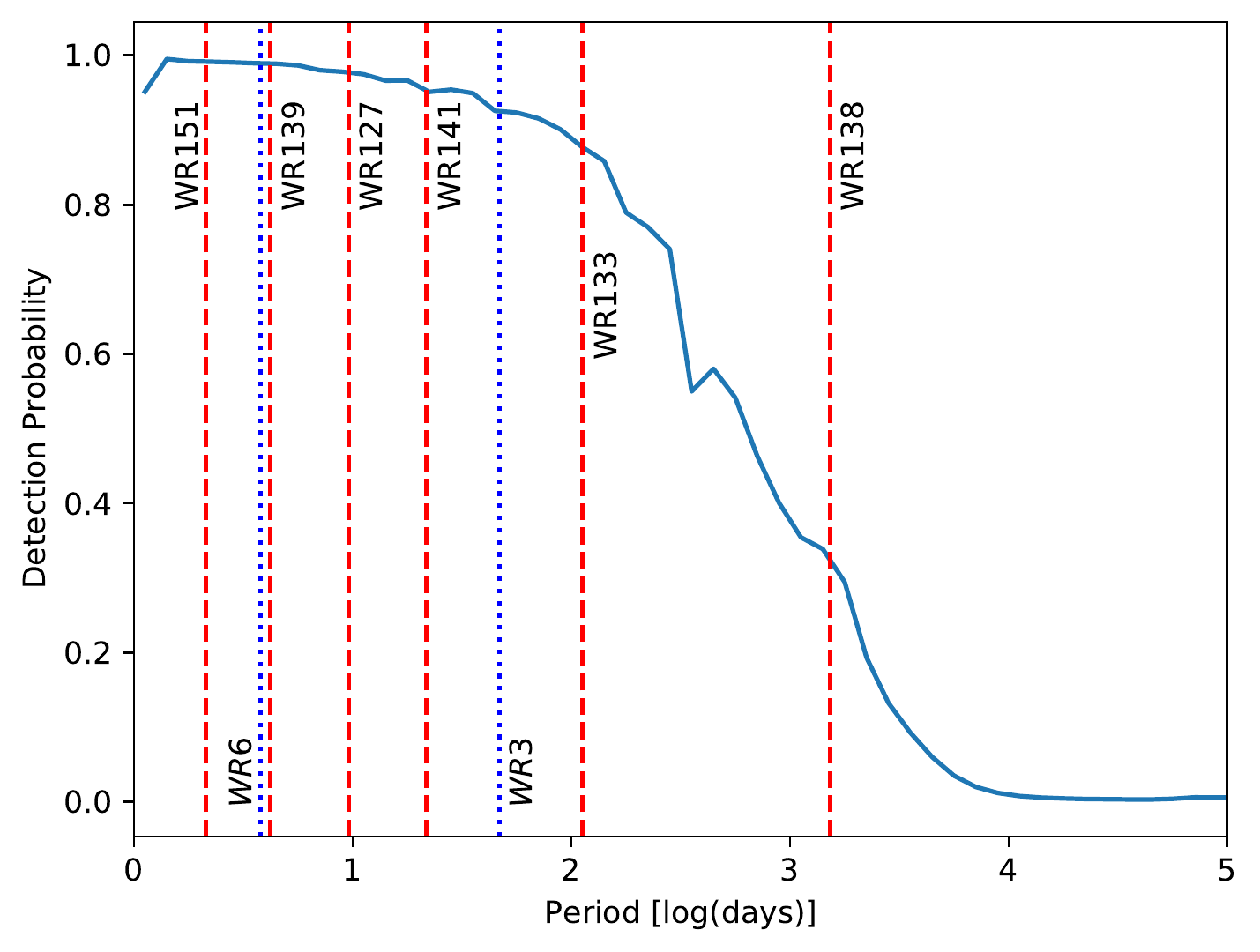}
    \caption{Binary detection probability of our campaign, assuming a flat mass-ratio distribution between 0.1 and 2.0 with a primary mass of 20\,\Msun{} and a flat eccentricity distribution between 0.0 and 0.9. The known binaries are marked with dashed red lines. WR 3 and WR 6 are marked with dashed blue lines because their reported periods cannot be verified with the available data.}
    \label{fig:Pdetect_overall}
\end{figure}

The RV time series at each grid point were simulated according to the observational sampling (number of epochs, time base) for each object. This allowed us to compute the probability that a given orbital configuration in $\log P$ and $M_2$ would give us an RV signal that would pass our detection threshold (i.e. \DelRV{}\,$\ge\,C=50$\,\kms{}). An example of such a probability detection plot is shown in Fig.  \ref{fig:Pdetect_WR1}. After we determined the probability detection plots for each object, we computed the average over the grid of $\log P$ and $M_2$. Assuming a flat mass-ratio between 0.1 and 2.0 with a primary mass of $M_1 = 20$\,\Msun{}, we averaged over the secondary mass to calculate the average detection probability as a function of orbital period (Fig. \ref{fig:Pdetect_overall}). In \citetalias{2020Dsilva}, we adopted $C=10\,$\kms{}, so that the efficiency of detecting WC binaries dropped below ${\sim}80$\% at orbital periods of 1000\,d (Fig. 8 in \citetalias{2020Dsilva}). Here, due to a larger adopted threshold $C$ for our WNE sample, this efficiency drops below 80\% already at periods of ${\sim}$100\,d (Fig.\ref{fig:Pdetect_WR1}). 

Figure \ref{fig:Pdetect_overall} shows the global detection probability of our RV campaign as a function of the orbital period. As a first approach, the observed binaries in our sample can be corrected with their respective detection probability, and thus the number of true binaries in our sample can be estimated. We ignored WR 3 in this exercise as we did not classify it as a RV variable system (Fig. \ref{fig:binfrac}), and we ignored WR 6 since we are unable to confirm the orbital period with our data (App. \ref{apdx:comments}). After defining a quantity $p(P)$ to be the detection probability for each binary detected with a given orbital period $P$, we expect our sample to contain 1/$p(P)$ binary systems. The largest correction factor comes from WR 138, which has a detection probability of ${\sim}0.33$. We therefore expect three similarly long period binaries (1/0.33) to be present in our sample on average. Given the six detected binaries in our sample, we expect about nine binaries in the sample, corresponding to an \fint{} of 0.56. This straightforward counting correction is only possible because the orbital periods of the binary systems are known, but it can be strongly affected by small sample statistics. 

\subsection{Adopting a Bayesian approach}\label{sect:bayesian}
Because $p(P)$ is a strong function of $P$, the bias-corrected binary fraction depends on the underlying period distribution. Here, we adopt a more formal approach to determine both within a Bayesian framework. The model parameters of a population with binaries include those that describe the period distribution, binary fraction, inclination, eccentricity, mass ratio, orientation of the orbital   plane, time of periastron passage, and so on. Ideally, we would like to calculate the likelihood of observing the RV time series for each object given these model parameters. However, this is computationally too intensive, and thus we adopted a different strategy. We divided the 16 WNE stars into four \DelRV{} bins: \DelRV{}\,$\le$\,50\,\kms{} (nine objects), 50\,$\le$\,\DelRV{}\,$\le$\,250\,\kms{} (four objects), 250\,$<$\,\DelRV{}\,$\le$\,650\,\kms{} (three objects) and  \DelRV{}\,$\ge$\,650\,\kms{} (no objects).

The shape of the period distribution is parametrised by its power-law index $\pi$ as follows:
\begin{equation}
    p(\log P) \sim (\log P)^\pi.
\end{equation}
The period distribution further has lower and upper bounds, called \logPmin{} and \logPmax{}. Along with \fint{}, we explored $\pi$, \logPmin{} , and \logPmax{} as model parameters. Because very short period WNE binaries are present in our sample, we fixed $\log P_\mathrm{min} = 0.15 $\,[d]. We explored \logPmax{} in the range  of 3.0 to 5.0 in steps of 0.1. We explored values of $\pi$ from $-1.0$ to $1.0$ in steps of 0.1, and \fint{} from 0.3 to 1.0 in steps of 0.02. For each bin in this three-dimensional parameter space, we simulated 40\,000 sets of 16 WNE stars, that is, over  $6\times 10^8$ populations. The assumptions for the eccentricity, inclination, and so on\, are the same as the MC simulations above (Sect. \ref{sect:detection_probability}). Based on our prior knowledge of the known orbital periods (Table~\ref{tab:WN_data}), when possible, we enforced that our simulations had at least seven binaries in period bins: $P > 1$\,d (seven systems), $P > 10$\,d (four systems), $P > 100$\,d (two systems), and $P >1000$\,d (one system).

\begin{figure*}[ht]
    \centering
    \includegraphics[width=0.85\textwidth]{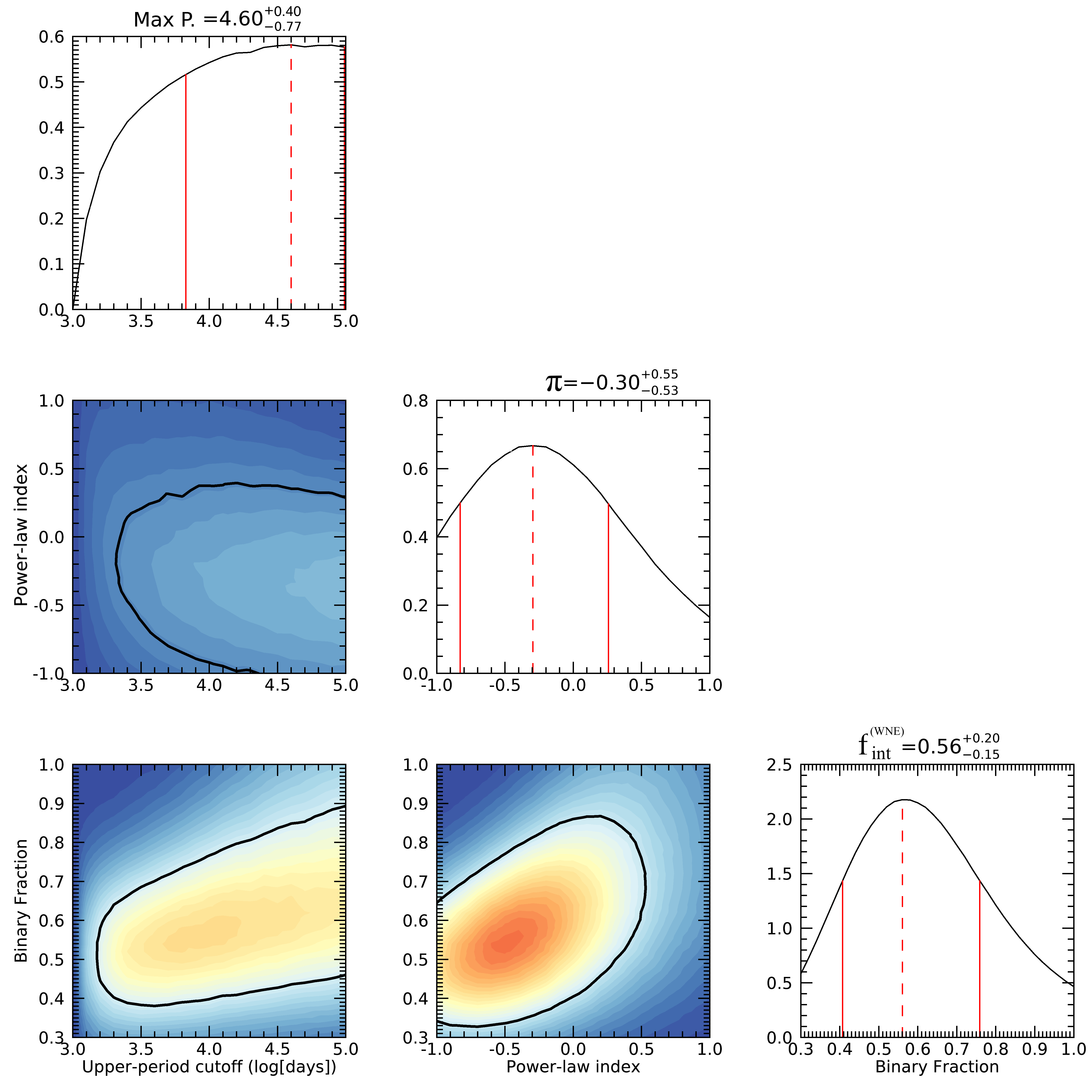}
    \caption{Three-dimensional likelihood over \logPmax{}, \fint{} , and $\pi$. Assuming flat priors, the one-dimensional posteriors are also shown. For each posterior, the solid red lines show HDI68, and the dashed red line shows the mode.}
    \label{fig:flat_pdist_binfrac}
\end{figure*}

We then proceeded to compute the three-dimensional likelihood of observing the different numbers of objects in \DelRV{} bins as a function of \logPmax{}, $\pi$ and \fint{} (Fig. \ref{fig:flat_pdist_binfrac}). Assuming flat priors for each of the three parameters, we computed the marginalised posterior likelihood for the individual parameters.  For each posterior, we report the mode and the 68\% highest-density interval (HDI68). Our computations yield HDI68s of \fint{}\,$= 0.56\substack{+0.20 \\ -0.15}$, \logPmax{}\,$= 4.60\substack{0.40 \\ -0.77}$ , and $\pi$\,$= -0.30\substack{+0.55 \\ -0.53}$. The upper bound of \logPmax{} is poorly constrained and is set by the maximum of the grid (5.00). However, an increase in \logPmax{} would simply result in an increase in \fint{} in order to maintain the fraction of binaries at shorter periods, which is required to reproduce our observations (Fig. \ref{fig:binfrac}). A negative value for $\pi$ indicates a slight preference for an overabundance of shorter-period systems. The obtained values are consistent with the values estimated for main-sequence O stars \citep[$f_{\textrm{bin}}=0.69\,\pm\,0.09$, $\pi=-0.55\,\pm\,0.22$:][]{2012Sana}. Similarly, our best-fit results suggest a long-period tail in the period distribution that extends to at least 20\,yr and probably more.

One of the caveats of our simulations is the assumption on the mass ratio distribution. We have assumed a flat distribution between 0.1 and 2. However, this need not be the case in nature, as the outcome of binary interaction scenarios is an important variable. Table \ref{tab:WN_data} shows that known companions around WNE stars are main-sequence O stars (i.e. more massive companions). However, this result may suffer from a bias because binaries with low-mass companions will be much harder to detect. This is because the WR companion will not show much RV variation, and the brightness ratio will make it hard to detect any spectral features of the companion. We therefore decided to explore various power-law indices for the mass ratio distribution ($\kappa$). With the same setup as above, we assumed $\kappa =$ +1.0 and -1.0 for the mass-ratio distribution and computed the same three-dimensional likelihood (Figs. \ref{fig:kappa_plus1} and \ref{fig:kappa_minus1}, respectively). The conclusions from Fig. \ref{fig:flat_pdist_binfrac} did not change significantly, with $\kappa = -1.0$ yielding more short-period binaries\,($\pi = -0.6$) alongside a larger \fint{} of 0.66, and $\kappa = +1.0$ yielding more massive companions with a slightly lower \fint{} of 0.54. In all cases, the constraint on the minimum extent of the period distribution remained similar with \logPmax{}\,$> 3.8$. 

\subsection{Revisiting the WC sample}
\label{sect:WC_bayesian}
We reanalysed our results from \citetalias{2020Dsilva} using the improved statistical framework presented in Sect.~\ref{sect:bayesian}. We calculated the four-dimensional likelihood and posteriors for \logPmin{}, \logPmax{}, \fint{} and $\pi$ in order to provide a fair comparison. Unlike in the WNE sample, we did not fix \Pmin{} because we do not have many observational constraints. We explored \logPmin{} from 0.15 up to values of 2.95 and \logPmax{} from 3.0 to 5.0, both in steps of 0.1. In order to fit the absence of short $P$ systems, a low number of intermediate $P$ and a large number of long $P$ binaries, we explored values of $\pi$ from $-1.0$ to $+4.0$ in steps of 0.1. We also explored values of \fintWC{} from 0.30 to 1.00 in steps of 0.02. As in Sects. \ref{sect:detection_probability} and \ref{sect:bayesian}, we assumed a flat distribution for the eccentricity between 0.0 and 0.9, random orientations of the orbital plane in three-dimensional space, and random times of periastron passage. 

\begin{figure*}[t]
    \centering
    \includegraphics[scale=0.8]{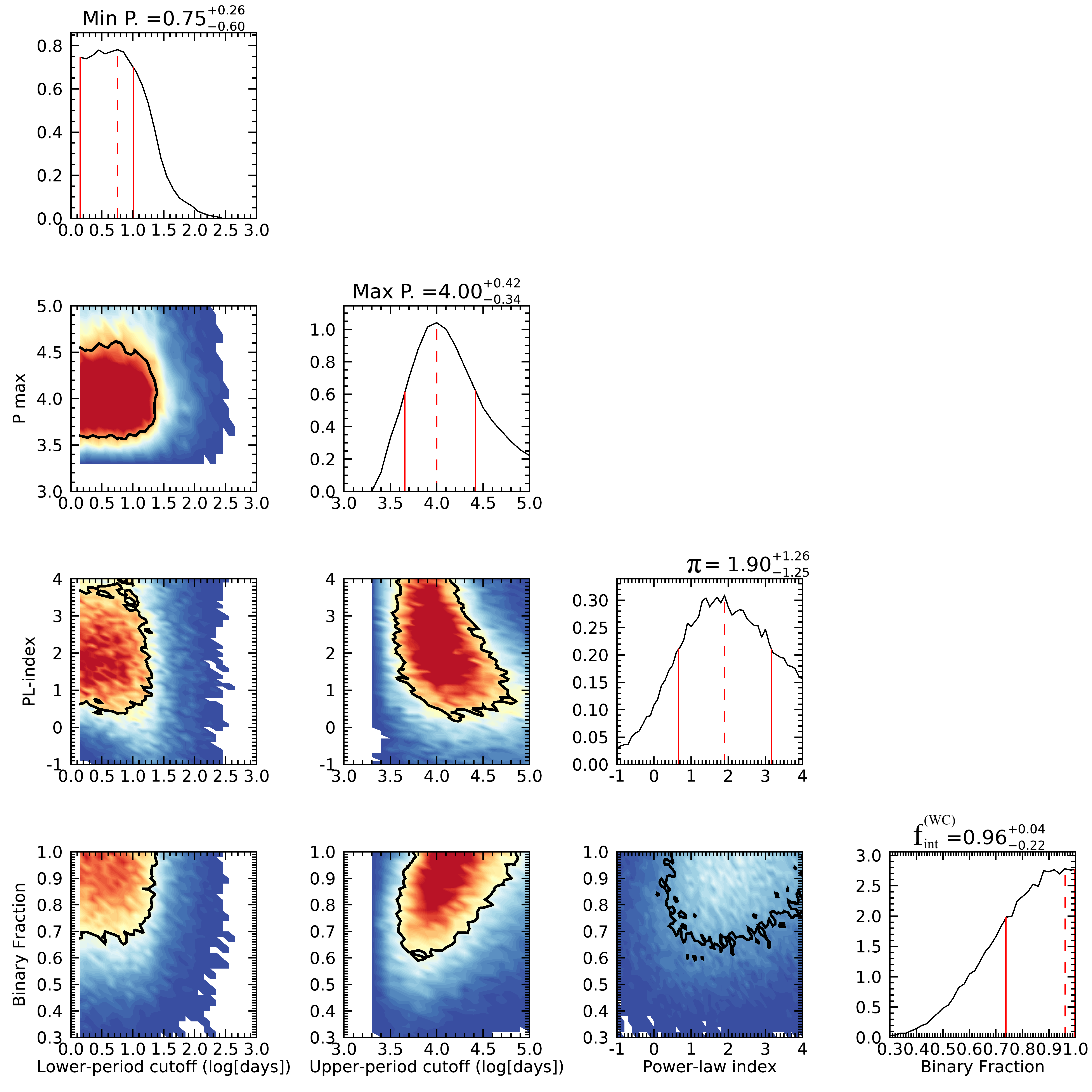}
    \caption{Four-dimensional likelihood over \logPmin{}, \logPmax{}, \fint{} , and $\pi$ for the WC sample. Assuming flat priors, the one-dimensional posteriors are also shown. For each posterior, the solid red lines show the HDI68, and the dashed red line shows the mode.}
    \label{fig:WC_posteriors}
\end{figure*}

The WC sample was separated into four \DelRV{} bins: 10\,$\le$\,\DelRV{}\,$\le$\,30\,\kms{} (six objects), 30\,$\le$\,\DelRV{}\,$\le$\,250\,\kms{} (no objects), 250\,$<$\,\DelRV{}\,$\le$\,300\,\kms{} (one object) and  \DelRV{}\,$\ge$\,300\,\kms{} (no objects). We then determined the likelihood of observing the same binned distribution for given values of the four model parameters. We also enforced that our simulations have binaries in the following period bins: $P > 20$\,d (three systems), $P > 2000$\,d (two systems). Assuming a primary mass of 10\,\Msun{}, we simulated 10\,000 populations of 12 WC stars for each bin in this four-dimensional parameter space, hence $\sim 1.1\times10^{10}$ populations. Fig.~\ref{fig:WC_posteriors} shows the four-dimensional likelihood plots along with the one-dimensional posteriors assuming flat priors. The white areas of the plot have a computed likelihood of zero, meaning that none of the 10\,000 populations computed for this set of parameters matched the observations. The HDI68 values are \fintWC{}\,$= 0.96\substack{+0.04 \\ -0.22}$, \logPmin{}\,$= 0.75\substack{+0.26 \\ -0.60}$, \logPmax{}\,$= 4.00\substack{+0.42 \\ -0.34}$ , and $\pi$\,$= 1.90\substack{+1.26 \\ -1.25}$. The values of \fintWC{} and $\pi$ are in agreement with the results of \citetalias{2020Dsilva}, where we found the majority of WC binaries with $P>100$\,d and \fintWC{}\,$> 0.72$.

Based on the above analyses, the visualisation of the intrinsic period distributions of the WC and WNE populations is presented in Fig. \ref{fig:periodDistShift}. The period distribution for the WNE population using the best-fit values of \logPmax{} and $\pi$, with \Pmin{} fixed at 1\,d, is shown in red. The solid blue line shows the period distribution for the WC population using the best-fit values of \logPmin{}, \logPmax{} , and $\pi$. In an attempt to express the uncertainties of these parameters on the period distribution, we sampled their posteriors 10\,000 times and over-plot the resulting constructed period distributions. The dark and light shaded areas show 68\% and 95\% of the highest density covered by the distributions, respectively. For the sake of comparison, the distribution for main-sequence O stars from \citet{2012Sana} is also shown (dashed black line). 

The derived multiplicity properties might be affected by spectral variability in WR stars on the RV measurements and by the relatively small sample of stars we used. While wind variability in WR stars has a significant effect on the measured RVs, we thoroughly quantified this using our high-cadence studies (Sect. \ref{sect:windVariability} and \citetalias{2020Dsilva}) and statistically accounted for it, using a variability threshold that is high enough to avoid false positives. Similarly, the Bayesian framework presented here accounts for the small sample size. Unless our sample is significantly biased and does not (in a statistical sense) reflect the properties of the parent WNE and WC populations, our conclusions that the WNE and WC stars have different period distributions should hold. We further elaborate on these two aspects in Sects. \ref{sect:literature} and \ref{sect:obsbias_brightness}.

%

\section{Discussion}\label{sect:evolution}

\subsection{Comparison of the binary properties with the literature} \label{sect:literature}

In order to assess whether our results might be biased by the small sample, we explored the literature for statistics on the WNE binary fraction. The GCWR has a total of 387 WN stars, of which 97 are classified as WNE\footnote{As indicated previously, we do not consider WN6 stars to be WNE stars.}. Most of the WR stars in the past decades have been detected through infrared surveys as they are embedded in dust and/or suffer from substantial reddening or extinction, and their multiplicity properties are poorly studied. In order to discuss \fobs{} in the context of a comparable sample that has been well studied for multiplicity, we considered the stars in \citetalias{2001vanderHucht}. The catalogue reports a total of 127 WN stars, of which 49 are of the WNE subtype. Of this subset of 49 stars, 13 are WNE binaries with derived spectroscopic orbital solutions. The observed spectroscopic multiplicity fraction for WNE stars in \citetalias{2001vanderHucht} is then 0.27. 

Table \ref{tab:WN_data} shows that most of
the WNE binary systems in our sample with known periods have periods of a few days to a hundred days. WR 138 is the only long-period binary system with a period of about four years. In \citetalias{2001vanderHucht}, all of the other WNE binaries with derived spectroscopic orbital solutions have orbital periods shorter than 15\,d. Therefore, the observed period distribution of the WNEs in our sample seems to be confirmed with the threefold larger literature-based sample. In contrast to the WC stars studied in \citetalias{2020Dsilva} and a similar literature search \citepalias{2001vanderHucht}, there seems to be a systematic skew in the observed period distribution, with a higher frequency of short-period WNE binaries than long-period ones, even in logarithmic space. There is thus no indication that our sample is biased, nor that the results of Sect. \ref{sect:intbinfrac} are contradicted by known WR properties from the existing literature at large.  
\subsection{Observational biases due to the magnitude limit} \label{sect:obsbias_brightness}

Considering a population of Galactic WR stars within a certain volume, it is reasonable to assume that the WR stars in binaries will be brighter than the single WR stars \citep{1980VanbeverenConti}. This implies a potential bias due to our selection criterion, which will be larger or smaller depending on the contrast in brightness between the binaries and the single stars.
Because most known WR binaries comprise OB-type companions, we made the simplifying assumption that WR binaries are twice as bright as single WR stars on average \citep[e.g.][]{2019Shenar}. This implies that to avoid this bias, we should include single WR stars in the $V$-band magnitude range 12.0 to 12.7 mag.

In order to investigate this, we explored the GCWR for WNE stars within this magnitude range. For stars with missing $V$-band magnitudes, we searched for $\varv$-band magnitudes between 13.0 and 13.7, similar to our approach in Sect. \ref{sect:sample}. For the WNE sample, we found two entries, both presumably single stars, WR 126 (WC5/WN) and WR 129 (WN4o). When we conservatively adopt these two targets as single WNE stars, \fint{} changes from 0.56 (9 out of 16) to 0.50 (9 out of 18). This difference is smaller by a factor of three than our errors and is hence negligible.

For the WC sample, we found four new entries in the GCWR. Of these, WR 114 \citep[WC5\,+\,OB?:][]{2001vanderHucht} and WR 132 \citep[WC6\,+\,?:][]{1983BisiacchiWR132} are candidate binaries, WR 125 \citep[WC7ed\,+\,O9III:][]{2019WilliamsNEOWISE,2021AroraWR125} is a confirmed binary, and WR 150 (WC5) is a presumably single star. As was the case with the WNEs, accounting for these objects results in \fintWC{} that is well within the derived uncertainties. 

Therefore, the effect of this bias is not significant given our derived values. This can be understood through multiple reasons. First, there is a void of non-extinct massive star-forming regions at distances between about 3 and 8\,kpc. While we may be able to detect WRs in the Carina star-forming region (${\sim}$2.7\,kpc), we are unable to detect them in Westerlund 1 (${\sim}$3.8\,kpc) because they suffer from large interstellar extinction \citep{2005Clark}. Second, a notable fraction of WC binaries are dust-producing systems \citep{1995Williamsdust} and hence could experience increased circumstellar reddening. This would cause them to be fainter in the $V$ band, which could balance out the increase in brightness due to additional companions.

\subsection{Implications for binary evolution}
According to our current understanding of stellar evolution \citep{1976Conti,2003MeynetMaeder,2007Crowther}, depending on their initial mass, main-sequence O stars evolve into either cool supergiants or luminous blue variables (LBVs) before becoming WR stars by losing their envelope. Because the spectra of WR stars are thought to reflect the products of fusion in the stellar interiors, the Conti scenario \citep{1976Conti,2007Crowther} proposes a spectroscopic evolution from O stars to the WNL, WNE, WC, and, if they are massive enough, WO phases before these stars end their lives as compact objects. The focus of this section is to use our new multiplicity constraints on the WC \citepalias{2020Dsilva} and WNE stars to investigate the evolutionary connections between these various categories of objects.


\begin{figure}
    \centering
    \includegraphics[width=0.49\textwidth]{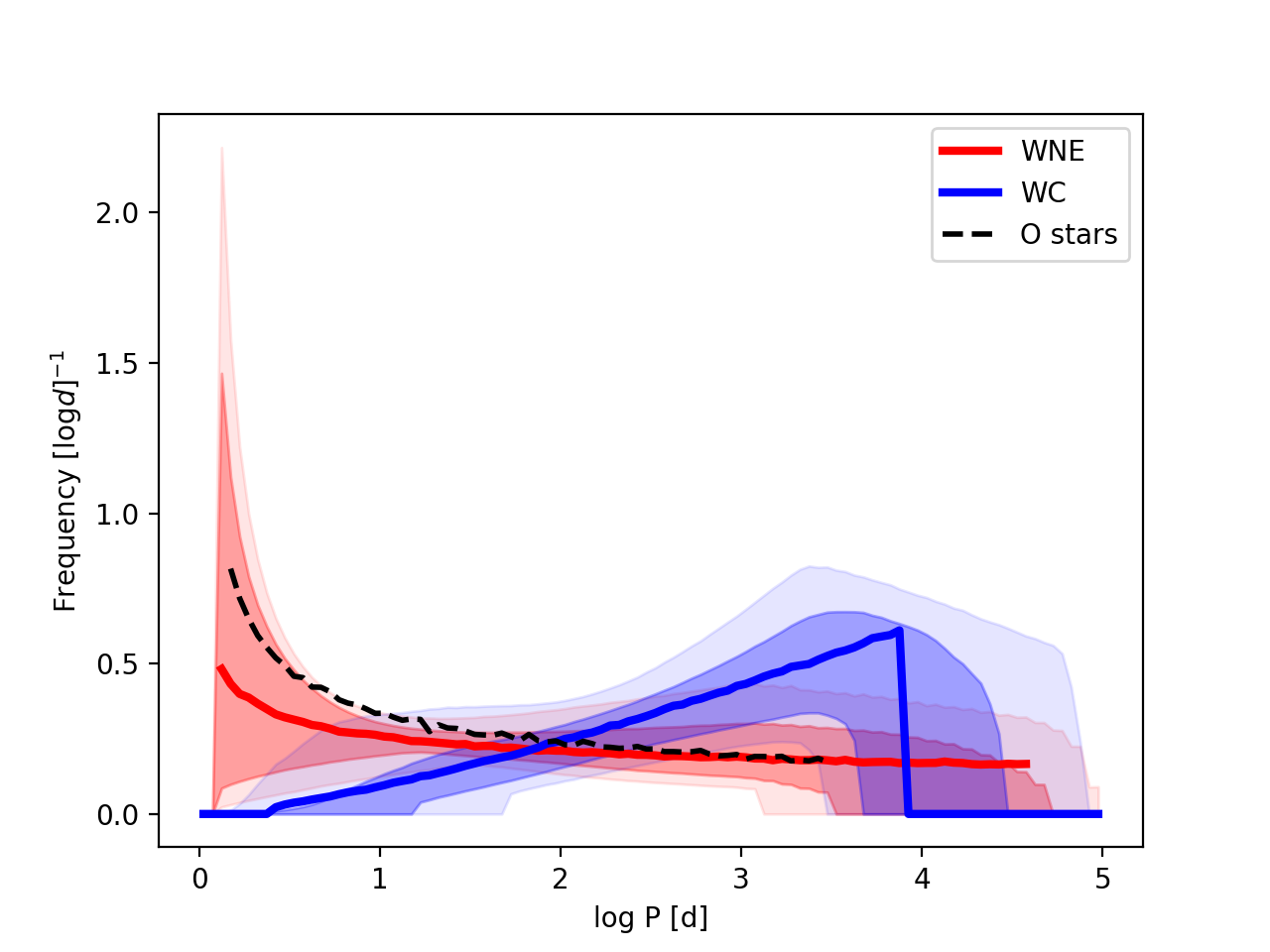}
    \caption{Period distributions for main-sequence O stars (dashed black line, from \citet{2012Sana}), WNE stars (red), and WC stars (blue). The WNE and WC distributions are calculated from the posteriors of the MC simulations (Sect. \ref{sect:bayesian}). The posteriors for the WC sample are shown in Fig. \ref{fig:WC_posteriors}.}
    \label{fig:periodDistShift}
\end{figure}

Following the Conti scenario, WNE stars are thought to have evolved from main-sequence O stars via the WNL phase by wind stripping. As discussed before, the multiplicity parameters of the populations of WNE and O stars are congruous within errors (Sect. \ref{sect:bayesian} and Fig. \ref{fig:periodDistShift}). If the population of binary WNE stars is the product of mass stripping in OB star binaries, similarities between their distributions are to be expected. For example, if a 40\,\Msun{} star becomes a 20\,\Msun{} WNE after transferring its envelope to a companion, the ratio of the pre- and post-interaction period ($P_\mathrm{f}/P_\mathrm{i}$) can be determined analytically given assumptions on mass-transfer efficiency and its companion \citep{1997SobermanMassTransfer}. For the case of a 20\,\Msun{} accretor and conservative mass transfer, $P_\mathrm{f}/P_\mathrm{i}=1$, while for fully non-conservative mass transfer (assuming material is ejected with the specific angular momentum of the accretor), we have $P_\mathrm{f}/P_\mathrm{i}\simeq 0.9$. As an example of a system with an initial mass ratio closer to unity, if the accretor has 35\,\Msun{} at the onset of mass transfer, then $P_\mathrm{f}/P_\mathrm{i}\simeq 2.1$ for conservative mass transfer, while $P_\mathrm{f}/P_\mathrm{i}\simeq 2.7$ for the non-conservative case. Therefore, mass transfer will not significantly modify the period distribution of O stars as they evolve into WNEs, which is compatible with the results of Fig. \ref{fig:periodDistShift}.

If WC stars are direct descendants of WNE stars, the orbital evolution from the WNE to the WC phase is expected to be mainly governed by mass loss due to their stellar winds. This results in the widening of the orbit, so that a shift to longer orbital periods is expected. To quantify this shift, we considered a 20\,\Msun{} WNE star in a binary system with a 30\,\Msun{} main-sequence O star. Even if the WNE star loses 15\,\Msun{} by the time it becomes a WC star, the orbital period of the system would only change by a factor of two. Therefore, only for WNE binaries in the long-period tail ($P>500$\,d) of the inferred distribution can mass loss lead to orbital periods compatible with the peak of the WC distribution ($P>1000$\,d). If indeed WC stars originate from WNE stars losing mass, then explaining the distributions shown in Fig. \ref{fig:periodDistShift} requires that preferentially longer-period WNE binaries evolve into WC stars, while the shorter-period ones avoid that outcome.

A possibility for the longer-period regime is that we simply do not detect WNE binaries with periods greater than a few thousand days in this current RV campaign. This is shown in Fig. \ref{fig:Pdetect_overall}, where our detection probability drops to 40\% at $P\sim1000$\,d and 0\% at $P>10000$\,d. The presence of an undetected long-period population of WNE binaries would indeed provide natural progenitors to the long-period WC binary population. The value of \fint{} would then be much higher than reported here, however. Almost all of the apparently single WNE stars would be in long-period binaries, resulting in a \fint{} close to 1.00.

\section{Conclusions}\label{sect:conclusions}
We have established the observed and intrinsic multiplicity properties of a complete magnitude-limited sample of 16 northern Galactic WNE stars using data from the HERMES spectrograph since 2017. We measured the RVs using cross-correlation with a statistical framework that allowed us to derive meaningful uncertainties. Adopting a peak-to-peak variability threshold $C$ of 50\,\kms{}, which is more than three times the observed short-term variability for WR 138, we derived an observed spectroscopic binary fraction, \fobs{} , of 0.44\,$\pm$\,0.12.

Using MC simulations, we determined constraints on a parametrised model of the distribution of orbital parameters. In particular, we considered a distribution described by a power-law index $\pi$ and upper and minimum values \logPmin{} and \logPmax{} for the period distribution, as well as the intrinsic binary fraction \fint{}. Assuming flat priors for these model parameters, we found \fint{} to be $0.56\substack{+0.20 \\ -0.15}$. The power-law index for the period distribution, $\pi$, was found to be $-0.30\substack{+0.55 \\ -0.53}$. Both these values are consistent with what was derived for main-sequence O stars by \citet{2012Sana}. Furthermore, the period distribution also favours the majority of systems at orbital periods shorter than 100\,d. This is in stark contrast to what we found for the Galactic WC population (\fintWC{}\,$= 0.96\substack{+0.04 \\ -0.22}$, $\pi$\,$= 1.90\substack{+1.26 \\ -1.25}$.), where the majority of binaries had orbital periods of a few 1000 days. The 68\% highest-density interval values of $\pi$ for the WC and WNE populations do not overlap.

Taking the discrepancy in the period distributions of the WNE and WC populations at face value in our analyses and in the literature \citepalias[][]{2001vanderHucht} questions the presence of a systematic evolutionary connection between WNE and WC stars. A population of WNE binaries at $P\,{\sim}\,1000$\,d appears to be missing. These would be ideal progenitors of the observed WC binaries given the orbital evolution due to mass loss. Secondly, the evolved counterparts of the observed short-period WNE binaries seem to be missing, despite the high sensitivity of the RV campaign in \citetalias{2020Dsilva}. These results, combined with a similar deficit from literature studies, indicate the absence of a corresponding population of short-period WC binaries.

Considering the populations of main-sequence O stars, WNE and WC stars are linked from an evolutionary standpoint, it is possible for main-sequence O binaries to evolve into WNE binaries through Case A or early Case B mass transfer. However, depending on the initial period, the system would not survive if the initial mass ratio diverged drastically from unity. It is also possible for short-period WNE binaries to form via a common-envelope scenario, which could occur due to either late Case B or Case C mass transfer in a wide main-sequence O binary, regardless of the initial mass ratio. The long-period tail of the WNE period distribution is consistent with the current multiplicity statistics for LBVs \citep{2022Mahy}, possibly indicating an evolutionary sequence from long-period main-sequence O binaries $\rightarrow$ LBV $\rightarrow$ WNE $\rightarrow$ WC.

The orbital evolution of WNE binaries is mainly governed by mass loss, leading to an expansion of the orbital period by a factor of ${\sim}1.5$-2. This change is insufficient to explain the observed differences between the WNE and WC period distributions, which peak at $P<10$\,d and $P{\sim}5000$\,d, respectively. It thus appears that short-period WN binaries have a low chance of becoming WC binaries, for reasons that are not yet understood.

The existing multiplicity properties gathered from the literature on a larger sample and the smaller but statistically better-characterised sample presented here seem to indicate that the underlying period distributions for WC and WNE populations are different. This can offer substantial diagnostics on the evolution of these systems. However, owing to the significant evolutionary implications from the above discrepancy, it is critical to increase the sample size of such magnitude-limited studies.


\begin{acknowledgements}

This work was published with funds from the European Research Council under European Union's Horizon 2020 research programme (grant agreement No. 772225). TS also acknowledges funding received from the European Union's Horizon 2020 under the Marie Skłodowska-Curie grant agreement No. 101024605. PM acknowledges support from the FWO junior postdoctoral fellowship No. 12ZY520N.
\end{acknowledgements}

%
%

\bibliographystyle{aa} 
\bibliography{references} 

\appendix
\section{Comments on specific objects}\label{apdx:comments}

\textbf{WR 1:}
According to the GCWR, WR 1 is classified as an SB1?. \citetads{1998MarchenkoMoffatPhotometry} found a variety of photometric periods with Hipparcos, but settled on the `best' value of 11.68\,$\pm$\,0.14\,d. As a follow-up, \citetads{1999aMorel} studied the variability of the different line profiles and measured centroid velocities of \HeII{}. They concluded that a non-degenerate companion causing this line-profile variability on short timescales would have to be luminous enough to be seen in the spectrum (either in spectral lines or through line dilution, which is not the case). They were unable to rule out the possibility of a compact object as a companion. Even though the X-ray emission from WR 1 is variable \citepads[]{1996Wessolowski}, the amount of X-rays is normal for its spectral type. 

Extensive studies of the line-profile variability were undertaken by \citetads{2007Flores} and \citetads{2010CheneStLouis}, who found periods of 7.684\,d and 16.9$^{+0.6}_{-0.3}$\,d, respectively. They both concluded that the periodicity was intrinsic to the stellar wind and not due to orbital motion. \citetads{2010CheneStLouis} proposed corotating interaction regions (CIRs) as the best scenario to explain their photometric and spectroscopic results. This was further supported by a spectropolarimetric study by \citetads{2013StLouis}, who found continuum polarisation at the level of 0.5\% for WR 1, indicating a large-scale structure in the wind. In this work, we find WR 1 to have a $\Delta$RV of 37.5\,\kms{} over ${\sim}1265$\,d, which is significantly below the threshold of 50\,\kms{} , and hence we classify it as a presumably single star.
\newline
\newline
\textbf{WR 2}: 
WR 2 is the only WN2 star in the Galaxy. It is classified as a visual binary in the GCWR and \citetads{2001vanderHucht}. The optical spectrum clearly shows absorption lines of a B star in it. With the help of optical spectroscopy, \citetads{2019Chene} did not find any significant RV variations over a period of ${\sim}10$\,yr. Using direct imaging with assumptions on the stellar parameters, the authors showed that the contribution of the B star to the optical spectrum is smaller than expected by a factor of ten (5\% instead of 50\% for its distance and absolute magnitude) and is hence likely to be a background object. We measure a $\Delta$RV of 36.9\,\kms{} for WR 2 over ${\sim}1085$\,d and classify it as a single star. 
\newline
\newline
\textbf{WR 3:} According to \citetads{2001vanderHucht}, WR 3 is classified as a WN3\,+\,O4 system. \citetads{1986Moffat+abs} found a period of 46.85\,$\pm$\,0.02\,d with a mean semi-amplitude of the RV curve, $K$, of 33\,\kms{}, using both He\,II 4686 (emission) and H$\gamma$ (absorption) lines. However, the RV curves of the absorption and the emission line were found to be roughly in phase (relative phase shift of 0.15\,$\pm$\,0.03), implying that the absorption lines are intrinsic to the WR star. Given that the line profiles of WR 3 are fairly stable and triangular, \citetads{2004MarchenkoMoffatCrowtherWR3} hypothesised it to be a single WN3 star with a high hydrogen abundance and were able to model the optical spectrum. The GCWR classifies it as a WN3ha (hydrogen rich with absorption lines).  In this work, we find WR 3 to have a $\Delta$ RV of 13.1\,\kms{} over ${\sim}$\,120\,d measured using the N\,V line at 4945\,\r{A}. We therefore reject the period found by \citetads{1986Moffat+abs} and classify it as a single star. 
\newline 
\newline
\textbf{WR 6:} The GCWR classifies WR 6 as a SB1? system. The system demonstrates spectacular photometric \citepads[][and references therein]{1998MarchenkoMoffatPhotometry} and spectral \citepads[]{1997Morel,2007Flores} variability with a period of 3.77\,d. Additionally, this periodicity has been shown to vary on longer timescales \citepads[][]{1989DrissenWR6,1992RobertWR6}. The line-profile variability was thought to be caused by wind-wind collisions with a (compact) binary companion, but \citetads{1997Morel} argued that this is rather due to a structured wind. 

\begin{figure}
    \centering
    \includegraphics[scale=0.6]{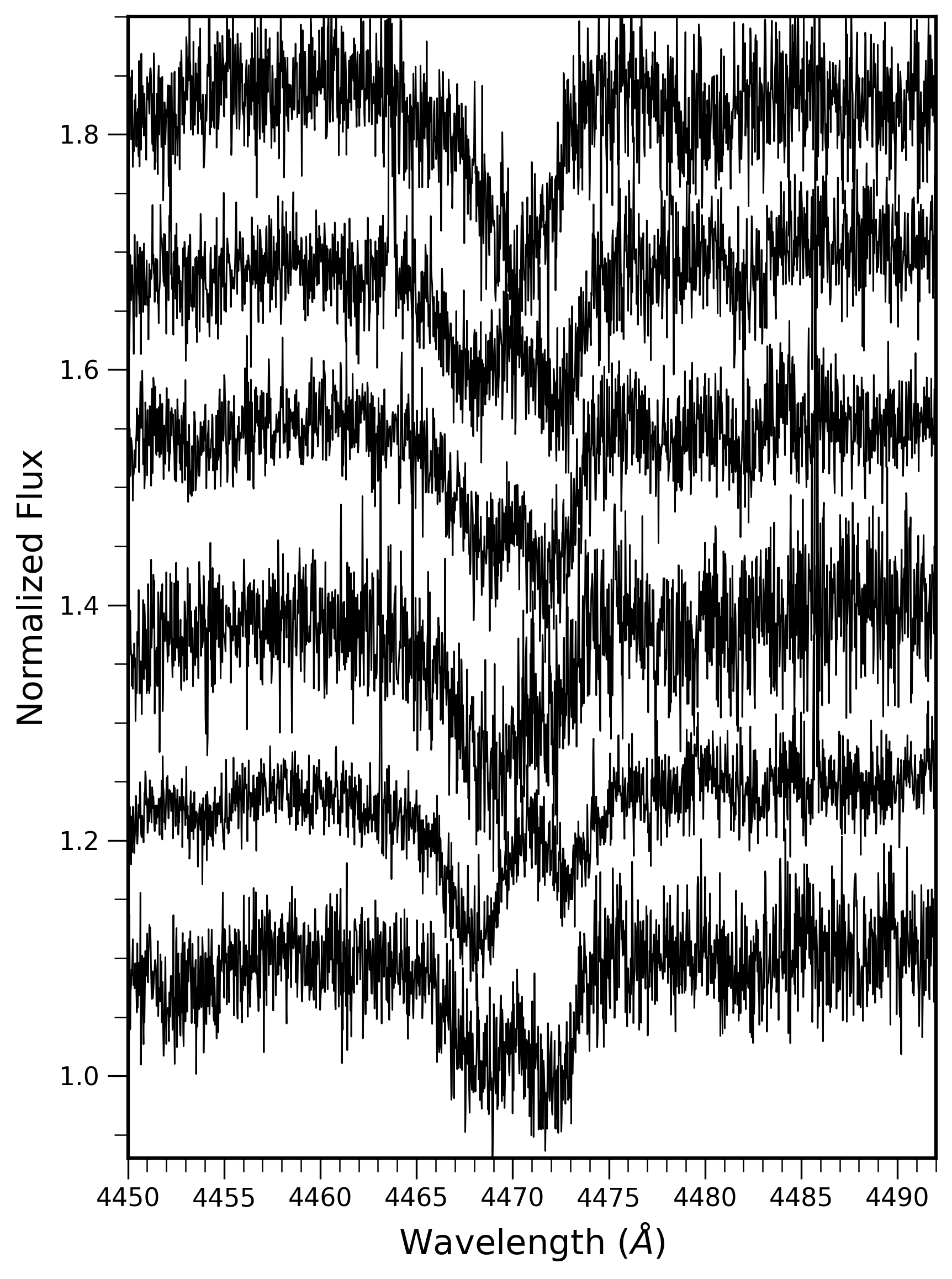}
    \caption{He\,{\sc{i}}\,$\lambda 4471$ line for WR 157. Six epochs of spectra are stacked with an arbitrary distance apart to demonstrate its SB3 nature.}
    \label{fig:WR157_he1}
\end{figure}

With uninterrupted photometry over 136\,d from the BRITE satellite, \citetads{2019SchmutzKoenigsbergerWR6} modelled the light curve and showed that the epoch-to-epoch variability was due to rapid apsidal motion in a binary. \citetads{2020Koenigsberger} used archival UV and X-ray data collected over several decades to model the emission peaks of various line profiles and also computed RVs. They were able to explain both the RVs and light curves with rapid apsidal motion in an eccentric binary (WR+B, e=0.1). Moreover, the authors also presented arguments for a third companion. In this work, we measure a $\Delta$RV of 170.0\,\kms{} over ${\sim}$1025\,d and classify WR 6 to be a binary system, in agreement with other findings. 
\newline
\newline
\textbf{WR 7:} The GCWR classifies WR 7 as a single star. With Hipparcos photometry, \citetads{1998MarchenkoMoffatPhotometry} observed a long-term trend but did not find any periods. We find that WR 7 exhibits strong line-profile variability over short timescales, although we were unable to discern any conclusive periodicity. With observations over ${\sim}$640\,d, we measure a $\Delta$RV of 46.5\,\kms{} and classify it as a single star. However, recent observations from TESS indicate the presence of two short periods, about 0.3\,d and 2.6\,d. A detailed analysis of these periods is in progress (Toal\'{a} et al. in prep).
\newline
\newline
\textbf{WR 10:} In the GCWR, WR 10 is classified as a single WN5h star. It is a visual binary with an A2V star as a companion, but it lacks RV variation \citepads[][]{1999aNiemela}, which we confirm. In this work, we find WR 10 to have a $\Delta$RV of 10.2\,\kms{} over ${\sim}$680\,d, thus classifying it as a single star. 
\newline
\newline
\textbf{WR 110:} WR 110 is classified as a single WN5b (broad-lined) star in the GCWR. Over a time span of ${\sim}$680\,d, we do not find significant RV variations with a $\Delta$RV of 31.8\,\kms{} and classify it as a single star. 
\newline
\newline
\textbf{WR 127:} According to the GCWR, WR 127 is a binary system with spectral type WN3b\,+\,O9.5V. A spectroscopic orbit was derived by \citetads{1981MasseyOrbits} with a period of 9.5550\,$\pm$\,0.0002\,d. \citetads{1996Lamontagne} and \citetads{1998MarchenkoMoffatPhotometry} found shallow eclipses in the light curve over the same period. \citet{2011delachevrotiere} improved the orbital solution, modelled the wind-wind collision zone, and revised the classification of the O-star, reclassifying WR 127 as a WN5o+O8.5V system. We find WR 127 to have a \DelRV{} of 332.0\,\kms{} over ${\sim}$2650\,d and classify it as a binary system.
\newline
\newline
\textbf{WR 128:} According to the GCWR, WR 128 is classified as a WN4(h) star with a binary status of SB2?. It was found to be a photometric variable by \citet{1985AntokhinCherepashchuk} with a period of 3.871\,d in the V band. The depth of the eclipses led the authors to conclude that the companion was a neutron star with a bright accretion disk. \citet{1986MoffatShara} could not verify this period in their study, and the nature of the companion has also not been confirmed yet. In this study, we find WR 128 to have a $\Delta$RV of 9.7\,\kms{} over ${\sim}$1120\,d and classify it as a single star. 
\newline
\newline
\textbf{WR 133:} WR 133 is classified as a WN5o\,+\,O9I system. In addition to being a SB2 system, it also shows photometric \citep{1998MarchenkoMoffatPhotometry} and polarimetric \citep{1989ARobert} variability, although the photometric variability is not coherent with the binary orbit. The spectroscopic orbit was first derived by \citet{1994UnderhillHill}. \citet{2021Richardson} combined interferometric and spectroscopic data to derive the first visual orbit for a WN star and also to improve upon the parameters of the system. In this study, we measure RVs for WR 133 with \NVred. We find a $\Delta$RV of 111.6\,\kms{} over a timescale of ${\sim}$2900\,d and classify it as a binary system.
\newline
\newline
\textbf{WR 138:} According to the GCWR, WR 138 is classified as a WN5o\,+\,B? system. The companion was classified as a O9 star by \citet{1990Annuk}, who found a period of 1538\,d. This was further confirmed and improved upon by \citet{2013Palate}, who derived a period of 1521\,$\pm$\,35\,d with optical spectroscopy. They also found evidence of wind-wind collision with X-ray observations. Finally, \citet{2016Richardson} were able to resolve the system with interferometry and, with spectroscopy, they classified the companion as an O9V star with a fairly high $\varv\,$sin\,$i$ of ${\sim}$350\,\kms{}. We measured the RVs with weak \nv{} lines, particularly a combination of \NVblue,{} and \NVred{}. Over a timescale of ${\sim}$2900\,d, we find WR 138 to have a $\Delta$RV of 92.1\,\kms{} and classify it as a binary system. 
\newline
\newline
\textbf{WR 139:} Also known as V444 Cyg, WR 139 is an eclipsing short-period binary system with a spectral classification of WN5o\,+\,O6III-V with a period of 4.212435\,d. It is a well-studied object both with photometry \citep[and references within]{1986MoffatShara,1998MarchenkoMoffatPhotometry} and spectroscopy \citep{1994Marchenko}. \citet{1994Marchenko} also found evidence of a stable, short-period signal (P${\sim}$0.36\,d) from the \HeII{} and He\,{\sc{ii}}\,$\lambda 5412$ radial velocities that they attributed to pulsations. The RVs were measured with \NVred. With a $\Delta$RV of 618.3\,\kms{} over ${\sim}$2950\,d, we classify WR 139 as a binary system. 
\newline
\newline
\textbf{WR 141:} According to the GCWR, WR 141 is a SB2 binary system with a spectral type of WN5o\,+\,O5V-III and a period of 21.7\,d. Orbital solutions were derived by both \citet{1998Marchenko1998WR141} and \citet{1999Ivanov}, but we list the former in Table \ref{tab:WN_data} as it is an SB2 solution and as both derived orbital parameters are consistent within errors. In this study, we find WR 141 to have a $
\Delta$RV of 209.8\,\kms{} over ${\sim}$2040\,d  and classify it as a binary system. The RVs were measured with \NVred.
\newline
\newline
\textbf{WR 151:} WR 151 is another short-period eclipsing binary with a period of 2.12691\,d that is classified as a WN4o\,+\,O5V system in the GCWR. \citet{2009HuttonWR151} combined RV measurements from \citet{1993LewisWR151} and photometry to fit a simultaneous orbital solution. We measure RVs with \NVred, find a $\Delta$RV of 650\,\kms{} over ${\sim}$950\,d and classify it as a binary.
\newline
\newline
\textbf{WR 152:} The GCWR classifies WR 152 as a WN3(h). We measure RVs using \NVred{} and find a $\Delta$RV of 8.3\,\kms{} over ${\sim}$920\,d, thus classifying it as a single star. 
\newline
\newline
\textbf{WR 157:} According to the GCWR, WR 157 is a visual binary with a spectral type WN5o\,(+B1II). As part of a larger sample, \citet{2021MaizAppelaniz} observed the system with spectroscopy and reclassified the companion as a B0.7 II star. In this study, we measure RVs for the WR star with \NVblue{} and note that it barely moves. We measure a $\Delta$RV of 11.8\,\kms{} over ${\sim}$2170\,d. However, He\,{\sc{i}}\,$\lambda 4471$ shows an SB2 behaviour, implying that the system is a triple (Fig.\,\ref{fig:WR157_he1}). 
\heii\,$\lambda 4541$ also appears to show the same SB2 behaviour, although it is less certain due to the blend with the WR emission. We conclude that the system has a tight OB\,+\,OB binary and a WR star whose connection to the OB binary is yet to be established. This is similar to BAT99 126, the WR star in the Large Magellanic Cloud that was found to be a quadruple system \citep{2021Janssens}. 
\newline
\newline
\section{Relative RV measurements}\label{s:tables_RV}
Relative RVs for the objects in our sample. The reference epoch has a RV of 0.0\,\kms{}. We refrain from providing absolute RV measurements as this is highly method dependent, especially for WR stars. The Barycentric Julian Date (BJD) is given as the middle of the exposure. The average S/N is given in Table\,\ref{tab:wr_epochs}. Along with the measured RVs, we have indicated the measurement error, that is, the statistical uncertainty $\sigma_p$ (Eq. \ref{eq:sigmaRV}). 
\begin{table}[h!]
    \centering
    \caption{Journal of HERMES observations for WR 1. Mask used: wings of the \NVred{}}
    \begin{tabular}{ccc} \hline \hline
        BJD $-$ 2450000 (d) & Relative RV (\kms) & $\sigma_p$ (\kms) \\ \hline
        7966.6224 & 12.6 & 3.5 \\ 
        8090.5018 & 5.9 & 6.6 \\ 
        8102.3783 & $-$2.5 & 3.9 \\ 
        8131.3316 & 0.0 & 2.5 \\ 
        8713.6443 & 8.6 & 3.1 \\ 
        8714.5128 & 7.4 & 2.6 \\ 
        8717.6878 & $-$6.5 & 3.7 \\ 
        8774.6470 & $-$3.1 & 4.7 \\ 
        8779.4524 & $-$24.9 & 4.6 \\ 
        8789.5124 & $-$18.2 & 6.7 \\ 
        8796.5305 & $-$7.1 & 4.0 \\ 
        8817.3615 & 3.0 & 5.9 \\ 
        8826.5518 & $-$6.6 & 5.1 \\ 
        8856.3320 & $-$11.5 & 8.4 \\ 
        8860.3339 & $-$14.5 & 4.4 \\ 
        8878.3833 & $-$14.8 & 3.4 \\ 
        9081.6208 & $-$6.3 & 3.9 \\ 
        9082.6225 & 7.1 & 3.7 \\ 
        9146.5251 & $-$4.2 & 3.4 \\ 
        9152.4444 & $-$8.3 & 4.4 \\ 
        9153.4682 & $-$7.5 & 4.5 \\ 
        9154.4621 & 3.9 & 6.0 \\ 
        9155.5839 & $-$4.4 & 6.3 \\ 
        9156.5588 & $-$15.5 & 4.7 \\ 
        9170.4167 & $-$11.1 & 5.4 \\ 
        9206.3990 & $-$9.8 & 4.2 \\ 
        9230.3722 & $-$7.8 & 6.2 \\ \hline
    \end{tabular}
    \label{tab:WR1}
\end{table}

\begin{table}[h!]
    \centering
    \caption{Journal of HERMES observations for WR 2. Mask used: full spec.}
    \begin{tabular}{ccc} \hline \hline
        BJD $-$ 2450000 (d) & Relative RV (\kms) & $\sigma_p$ (\kms) \\ \hline
        8089.4747 & 13.7 & 1.4 \\ 
        8102.3978 & $-$23.2 & 4.8 \\ 
        8131.3613 & 0.0 & 3.4 \\ 
        8739.6347 & $-$16.1 & 4.3 \\ 
        8742.6421 & $-$10.4 & 4.5 \\ 
        8744.6089 & $-$13.6 & 4.9 \\ 
        8753.6487 & $-$7.1 & 3.3 \\ 
        8774.6743 & $-$8.9 & 3.5 \\ 
        8817.4228 & $-$16.8 & 5.4 \\ 
        9081.6507 & 2.8 & 3.2 \\ 
        9089.6989 & 8.9 & 3.0 \\ 
        9146.5668 & $-$1.2 & 3.4 \\ 
        9148.4564 & $-$4.4 & 5.6 \\ 
        9170.4916 & $-$15.0 & 4.2 \\ 
        9171.3888 & $-$5.4 & 4.3 \\ 
        9172.3591 & $-$2.3 & 4.7 \\ 
        9173.4578 & $-$9.6 & 4.4 \\ 
        9174.3348 & $-$0.4 & 3.6 \\ \hline
    \end{tabular}
    \label{tab:WR2}
\end{table}

\begin{table}[h!]
    \centering
    \caption{Journal of HERMES observations for WR 3. Mask used: \NVred{}.}
    \begin{tabular}{ccc} \hline \hline
        BJD $-$ 2450000 (d) & Relative RV (\kms) & $\sigma_p$ (\kms) \\ \hline
        8786.6329 & $-$3.4 & 0.9 \\ 
        8817.4783 & 0.6 & 1.1 \\ 
        8854.4345 & 0.0 & 0.9 \\ 
        8876.3900 & 0.1 & 1.0 \\ 
        8878.4478 & 1.5 & 1.1 \\ 
        8892.3619 & 1.1 & 1.5 \\ 
        8898.3471 & $-$0.7 & 1.3 \\ 
        8906.3982 & $-$3.5 & 1.1 \\ \hline
    \end{tabular}
    \label{tab:WR3}
\end{table}

\begin{table}[h!]
    \centering
    \caption{Journal of HERMES observations for WR 6. Mask used: full spec.}
    \begin{tabular}{ccc} \hline \hline
        BJD $-$ 2450000 (d) & Relative RV (\kms{}) & $\sigma_p$ (\kms{}) \\ \hline
        8131.5571 & 2.7 & 6.0 \\ 
        8139.5854 & 3.9 & 5.5 \\ 
        8165.4858 & $-$7.9 & 10.3 \\ 
        8220.3593 & $-$55.0 & 7.9 \\ 
        8227.3515 & 62.1 & 7.0 \\ 
        8796.7597 & 14.2 & 6.7 \\ 
        8799.7682 & $-$106.0 & 6.4 \\ 
        8804.7649 & 0.0 & 5.4 \\ 
        8809.6896 & $-$83.2 & 4.5 \\ 
        8818.6700 & $-$107.9 & 3.7 \\ 
        8854.5539 & $-$67.1 & 5.1 \\ 
        8859.5353 & $-$98.4 & 4.4 \\ 
        8876.4221 & $-$50.7 & 4.4 \\ 
        8876.4272 & $-$60.1 & 4.0 \\ 
        8876.4324 & $-$48.1 & 4.2 \\ 
        8876.5373 & $-$55.3 & 4.0 \\ 
        8876.5425 & $-$55.3 & 4.2 \\ 
        8876.5477 & $-$48.8 & 4.1 \\ 
        8876.5529 & $-$52.5 & 4.4 \\ 
        8876.5581 & $-$52.9 & 4.8 \\ 
        8876.5634 & $-$58.2 & 3.7 \\ 
        9147.7394 & $-$6.6 & 7.0 \\ 
        9149.7643 & $-$0.7 & 7.3 \\ 
        9150.7615 & $-$57.9 & 6.6 \\ 
        9151.7686 & $-$2.5 & 5.9 \\ 
        9152.7737 & 0.2 & 9.6 \\ 
        9155.7095 & 4.4 & 5.8 \\ 
        9156.7172 & $-$11.3 & 6.7 \\ \hline
    \end{tabular}
    \label{tab:WR6}
\end{table}

\begin{table}[h!]
    \centering
    \caption{Journal of HERMES observations for WR 7. Mask used: \NVred.}
    \begin{tabular}{ccc} \hline \hline
        BJD $-$ 2450000 (d) & Relative RV (\kms) & $\sigma_p$ (\kms) \\ \hline
        8167.4273 & $-$11.5 & 1.2 \\ 
        8183.3812 & $-$10.4 & 1.2 \\ 
        8214.4138 & 0.0 & 1.1 \\ 
        8220.3859 & $-$20.9 & 1.1 \\ 
        8225.3718 & $-$18.4 & 1.3 \\ 
        8786.7478 & $-$25.3 & 1.5 \\ 
        8809.7142 & $-$46.5 & 1.9 \\ \hline
    \end{tabular}
    \label{tab:WR7}
\end{table}

\begin{table}[h!]
    \centering
    \caption{Journal of HERMES observations for WR 10. Mask used: full spec.}
    \begin{tabular}{ccc} \hline \hline
        BJD $-$ 2450000 (d) & Relative RV (\kms) & $\sigma_p$ (\kms) \\ \hline
        8139.6049 & 2.7 & 0.6 \\ 
        8166.4471 & 0.0 & 0.7 \\ 
        8188.4025 & $-$0.7 & 0.7 \\ 
        8223.3767 & 3.4 & 0.8 \\ 
        8226.3695 & $-$4.5 & 0.6 \\ 
        8817.7041 & 5.7 & 0.8 \\ \hline
    \end{tabular}
    \label{tab:WR10}
\end{table}

\begin{table}[h!]
    \centering
    \caption{Journal of HERMES observations for WR 110. Mask used: full spec.}
    \begin{tabular}{ccc} \hline \hline
        BJD $-$ 2450000 (d) & Relative RV (\kms) & $\sigma_p$ (\kms) \\ \hline
        7880.6775 & $-$2.4 & 2.1 \\ 
        7904.6723 & 1.9 & 2.3 \\ 
        7907.6736 & $-$9.4 & 2.8 \\ 
        8227.7273 & $-$14.1 & 3.1 \\ 
        8249.6929 & $-$20.3 & 2.2 \\ 
        8255.6409 & 4.7 & 2.8 \\ 
        8662.5573 & 0.0 & 2.8 \\ 
        8717.4911 & $-$27.1 & 5.9 \\ 
        8721.4891 & $-$19.1 & 4.3 \\ 
        9002.6274 & 4.1 & 3.4 \\ 
        9022.5824 & $-$13.7 & 2.5 \\ \hline
    \end{tabular}
    \label{tab:WR110}
\end{table}

\begin{table}[h!]
    \centering
    \caption{Journal of HERMES observations for WR 127.  Mask used: \NVred.}
    \begin{tabular}{ccc} \hline \hline
        BJD $-$ 2450000 (d) & Relative RV (\kms) & $\sigma_p$ (\kms) \\ \hline
        6383.7311 & 269.4 & 2.5 \\ 
        7925.6481 & 190.6 & 2.7 \\ 
        7958.5570 & 128.5 & 6.3 \\ 
        7965.6804 & 27.0 & 3.2 \\ 
        8262.6484 & 0.0 & 2.9 \\ 
        8317.6849 & 137.1 & 3.0 \\ 
        8624.6814 & 36.4 & 5.2 \\ 
        8669.6351 & 292.6 & 3.3 \\ 
        8677.4986 & 332.0 & 3.4 \\ 
        9001.5189 & 274.0 & 2.2 \\ 
        9029.5713 & 201.6 & 2.5 \\ \hline
    \end{tabular}
    \label{tab:WR127}
\end{table}

\begin{table}[h!]
    \centering
    \caption{Journal of HERMES observations for WR 128. Mask used: \nv{} weak.}
    \begin{tabular}{ccc} \hline \hline
        BJD $-$ 2450000 (d) & Relative RV (\kms) & $\sigma_p$ (\kms) \\ \hline
        7914.7009 & $-$3.9 & 0.8 \\ 
        7949.6014 & $-$5.4 & 1.3 \\ 
        7965.6303 & $-$6.9 & 1.1 \\ 
        7965.6540 & $-$2.6 & 1.1 \\ 
        8201.7115 & $-$9.5 & 1.0 \\ 
        8678.6017 & $-$9.7 & 1.4 \\ 
        8749.5071 & 0.0 & 1.1 \\ 
        8762.4726 & $-$4.6 & 1.6 \\ 
        9009.6923 & $-$4.1 & 0.8 \\ 
        9034.6393 & $-$4.4 & 1.3 \\ \hline
    \end{tabular}
    \label{tab:WR128}
\end{table}

\begin{table}[h!]
    \centering
    \caption{Journal of HERMES observations for WR 133. Mask used: \NVred{}.}
    \begin{tabular}{ccc} \hline \hline
        BJD $-$ 2450000 (d) & Relative RV (\kms) & $\sigma_p$ (\kms) \\ \hline
        6118.7225 & 57.5 & 3.7 \\ 
        6118.7252 & 58.1 & 4.2 \\ 
        6118.7279 & 60.1 & 3.8 \\ 
        6488.7325 & 0.0 & 1.8 \\ 
        6889.5771 & 80.7 & 2.1 \\ 
        6890.5661 & 93.8 & 2.1 \\ 
        7896.6971 & 79.1 & 3.5 \\ 
        7902.6919 & 77.7 & 3.3 \\ 
        7914.7239 & 68.8 & 3.3 \\ 
        8184.7739 & 42.7 & 4.0 \\ 
        8206.6924 & 27.4 & 4.1 \\ 
        8262.6628 & 30.1 & 2.7 \\ 
        8308.6629 & 41.2 & 4.0 \\ 
        8310.7398 & 24.0 & 4.1 \\ 
        8608.7226 & 20.2 & 3.2 \\ 
        8626.7249 & $-$17.8 & 2.8 \\ 
        8652.5782 & 20.2 & 2.6 \\ 
        8779.4212 & 59.0 & 4.7 \\ 
        8921.7629 & 80.9 & 6.3 \\ 
        8999.6873 & 48.6 & 4.0 \\ 
        9008.7106 & 44.7 & 3.1 \\ \hline
    \end{tabular}
    \label{tab:WR133}
\end{table}

\begin{table}[h!]
    \centering
    \caption{Journal of HERMES observations for WR 138. Mask used: \nv{} weak.}
    \begin{tabular}{ccc} \hline \hline
        BJD $-$ 2450000 (d) & Relative RV (\kms) & $\sigma_p$ (\kms) \\ \hline
        6126.4561 & 87.3 & 2.0 \\ 
        6126.4602 & 81.2 & 1.9 \\ 
        6126.4643 & 75.8 & 2.0 \\ 
        6890.5817 & 29.4 & 0.8 \\ 
        7902.7238 & 83.0 & 1.1 \\ 
        7938.6839 & 82.0 & 1.1 \\ 
        7966.6459 & 73.3 & 1.3 \\ 
        8199.7580 & 57.1 & 1.4 \\ 
        8209.7583 & 48.5 & 1.3 \\ 
        8265.6425 & 48.4 & 1.3 \\ 
        8309.6314 & 42.2 & 1.8 \\ 
        8629.7285 & 14.5 & 1.3 \\ 
        8705.5768 & $-$3.1 & 1.7 \\ 
        8706.6016 & 5.1 & 1.2 \\ 
        8707.5636 & 6.0 & 1.7 \\ 
        8708.5046 & $-$2.7 & 1.2 \\ 
        8709.4356 & 4.9 & 1.2 \\ 
        8710.5668 & 0.0 & 1.3 \\ 
        8712.5364 & 4.4 & 1.6 \\ 
        8713.5060 & 5.7 & 1.6 \\ 
        8714.4872 & 6.9 & 1.4 \\ 
        8715.4997 & 9.0 & 1.3 \\ 
        8716.4250 & 5.5 & 1.3 \\ 
        8716.5228 & 5.2 & 1.2 \\ 
        8716.6573 & 3.3 & 1.4 \\ 
        8717.5872 & 9.3 & 1.5 \\ 
        8718.4849 & 5.0 & 1.3 \\ 
        8719.4380 & 3.8 & 1.5 \\ 
        8720.4098 & 4.8 & 2.2 \\ 
        8721.5385 & $-$4.9 & 1.5 \\ 
        8726.4055 & 10.4 & 1.2 \\ 
        8726.4439 & 1.0 & 1.5 \\ 
        8726.4945 & 5.3 & 1.3 \\ 
        8726.5372 & 4.2 & 1.3 \\ 
        8726.5798 & 7.5 & 1.4 \\ 
        8726.6220 & 7.4 & 1.3 \\ 
        8796.4390 & 7.7 & 1.6 \\ 
        8800.3992 & 2.3 & 2.2 \\ 
        9008.7211 & 41.9 & 1.2 \\ 
        9024.5965 & 37.9 & 1.2 \\ \hline
    \end{tabular}
    \label{tab:WR138}
\end{table}

\begin{table}[h!]
    \centering
    \caption{Journal of HERMES observations for WR 139. Mask used: \NVred{}.}
    \begin{tabular}{ccc} \hline \hline
        BJD $-$ 2450000 (d) & Relative RV (\kms) & $\sigma_p$ (\kms) \\ \hline
        6126.6048 & $-$217.3 & 8.1 \\ 
        6126.6090 & $-$232.2 & 9.0 \\ 
        6126.6131 & $-$231.8 & 7.4 \\ 
        6889.3780 & $-$80.4 & 1.4 \\ 
        6889.5173 & $-$38.7 & 1.4 \\ 
        6889.6903 & 29.5 & 3.0 \\ 
        6890.3932 & 274.2 & 1.3 \\ 
        6890.5345 & 280.2 & 1.2 \\ 
        6890.6132 & 275.1 & 2.8 \\ 
        6892.4109 & $-$290.3 & 4.3 \\ 
        6892.5832 & $-$338.1 & 5.7 \\ 
        6892.6850 & $-$285.8 & 6.1 \\ 
        6894.3711 & 212.3 & 1.3 \\ 
        6894.5183 & 261.5 & 1.0 \\ 
        6894.6484 & 259.0 & 1.2 \\ 
        6896.4061 & $-$236.0 & 3.2 \\ 
        6896.5401 & $-$247.7 & 6.9 \\ 
        6896.6318 & $-$274.3 & 5.3 \\ 
        6931.5018 & $-$100.1 & 2.2 \\ 
        6946.3462 & 0.0 & 2.1 \\ 
        6946.4447 & $-$64.2 & 2.8 \\ 
        7915.7128 & $-$227.3 & 6.0 \\ 
        7919.7200 & $-$130.5 & 3.8 \\ 
        7938.6750 & 78.2 & 1.9 \\ 
        8199.7629 & 21.2 & 2.7 \\ 
        8267.5991 & 192.4 & 2.7 \\ 
        8277.7066 & $-$95.6 & 3.0 \\ 
        8277.7119 & $-$91.1 & 3.6 \\ 
        8663.7305 & 247.9 & 1.9 \\ 
        8682.5103 & $-$243.0 & 5.7 \\ 
        8741.4968 & $-$250.7 & 5.2 \\ 
        8779.4428 & $-$281.8 & 11.3 \\ 
        9024.6078 & $-$255.7 & 8.2 \\ 
        9080.5344 & 167.5 & 2.2 \\ \hline
    \end{tabular}
    \label{tab:WR139}
\end{table}

\begin{table}[h!]
    \centering
    \caption{Journal of HERMES observations for WR 141. Mask used: \NVred.}
    \begin{tabular}{ccc} \hline \hline
        BJD $-$ 2450000 (d) & Relative RV (\kms) & $\sigma_p$ (\kms) \\ \hline
        6889.6476 & 74.8 & 5.1 \\ 
        6889.6690 & 47.7 & 7.9 \\ 
        7949.5642 & $-$11.9 & 4.8 \\ 
        7957.6105 & 126.0 & 4.3 \\ 
        7961.6127 & 59.2 & 4.6 \\ 
        8196.7557 & 188.7 & 3.0 \\ 
        8267.6131 & 0.0 & 2.7 \\ 
        8332.7166 & 14.9 & 2.9 \\ 
        8666.7123 & 46.6 & 3.4 \\ 
        8749.5352 & $-$15.0 & 3.0 \\ 
        8759.4257 & 152.6 & 4.4 \\ 
        8799.4244 & 136.5 & 7.4 \\ 
        9021.7199 & 146.7 & 6.8 \\ 
        9029.7067 & $-$21.1 & 6.9 \\ \hline
    \end{tabular}
    \label{tab:WR141}
\end{table}

\begin{table}[h!]
    \centering
    \caption{Journal of HERMES observations for WR 151. Mask used: \NVred.}
    \begin{tabular}{ccc} \hline \hline
        BJD $-$ 2450000 (d) & Relative RV (\kms) & $\sigma_p$ (\kms) \\ \hline
        8133.3587 & 0.0 & 9.5 \\ 
        8718.5857 & 124.8 & 19.3 \\ 
        8763.4920 & 351.8 & 31.2 \\ 
        9042.5982 & 650.0 & 41.3 \\ 
        9051.6960 & 277.9 & 15.9 \\ 
        9057.6630 & 622.4 & 34.0 \\ \hline
    \end{tabular}
    \label{tab:WR151}
\end{table}

\begin{table}[h!]
    \centering
    \caption{Journal of HERMES observations for WR 152. Mask used: \NVred.}
    \begin{tabular}{ccc} \hline \hline
        BJD $-$ 2450000 (d) & Relative RV (\kms) & $\sigma_p$ (\kms) \\ \hline
        8137.3704 & 0.0 & 1.2 \\ 
        8712.5811 & 4.7 & 1.5 \\ 
        8753.5905 & $-$0.0 & 1.6 \\ 
        8775.5911 & $-$3.0 & 1.5 \\ 
        8791.4069 & $-$2.0 & 1.3 \\ 
        9052.7004 & $-$3.6 & 1.7 \\ 
        9058.6666 & $-$0.6 & 1.5 \\ \hline
    \end{tabular}
    \label{tab:WR152}
\end{table}

\begin{table}[h!]
    \centering
    \caption{Journal of HERMES observations for WR 157. Mask used: \NVblue{}.}
    \begin{tabular}{ccc} \hline \hline
        BJD $-$ 2450000 (d) & Relative RV (\kms) & $\sigma_p$ (\kms) \\ \hline
        6890.6373 & $-$1.3 & 1.5 \\ 
        6890.6588 & $-$1.9 & 1.4 \\ 
        8091.4134 & 0.0 & 1.1 \\ 
        8321.7169 & $-$3.3 & 1.8 \\ 
        8336.6816 & $-$8.1 & 1.4 \\ 
        8665.6904 & $-$6.4 & 1.6 \\ 
        8754.5808 & $-$8.7 & 1.6 \\ 
        8762.5811 & $-$4.2 & 1.4 \\ 
        8775.4222 & 0.1 & 1.3 \\ 
        8787.4261 & 3.1 & 2.6 \\ 
        9041.7175 & $-$0.1 & 1.2 \\ 
        9058.7028 & $-$1.1 & 1.6 \\ \hline
    \end{tabular}
    \label{tab:WR157}
\end{table}

\section{Posterior plots from MC simulations}\label{s:posteriors}

\begin{figure}
    \centering
    \includegraphics[scale=0.4]{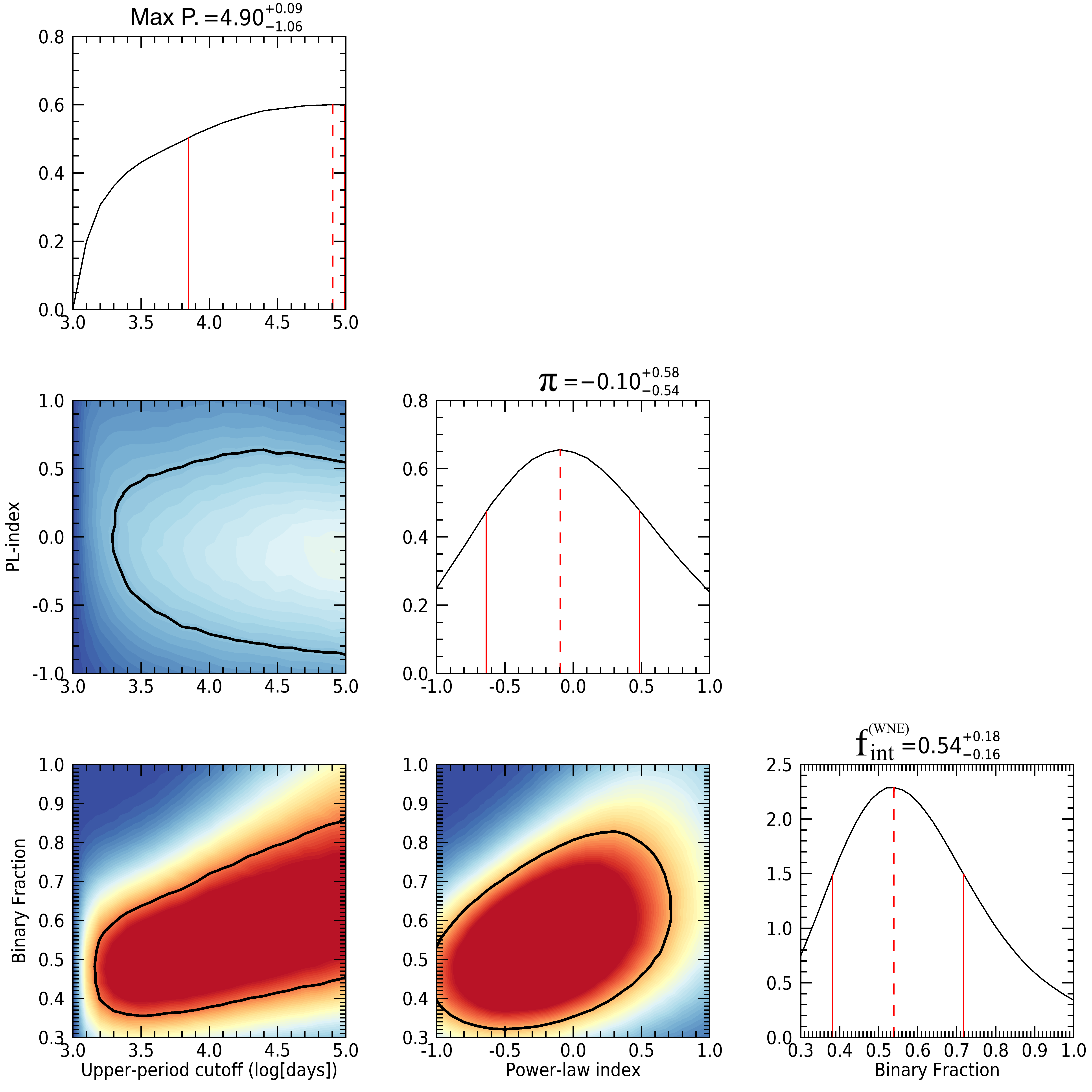}
    \caption{Posteriors for \logPmax{}, \fint{} , and $\pi$ for $\kappa = +1.0$. For each posterior, the solid red lines show HDI68, and the dashed red line shows the mode.}
    \label{fig:kappa_plus1}
\end{figure}

\begin{figure}
    \centering
    \includegraphics[scale=0.4]{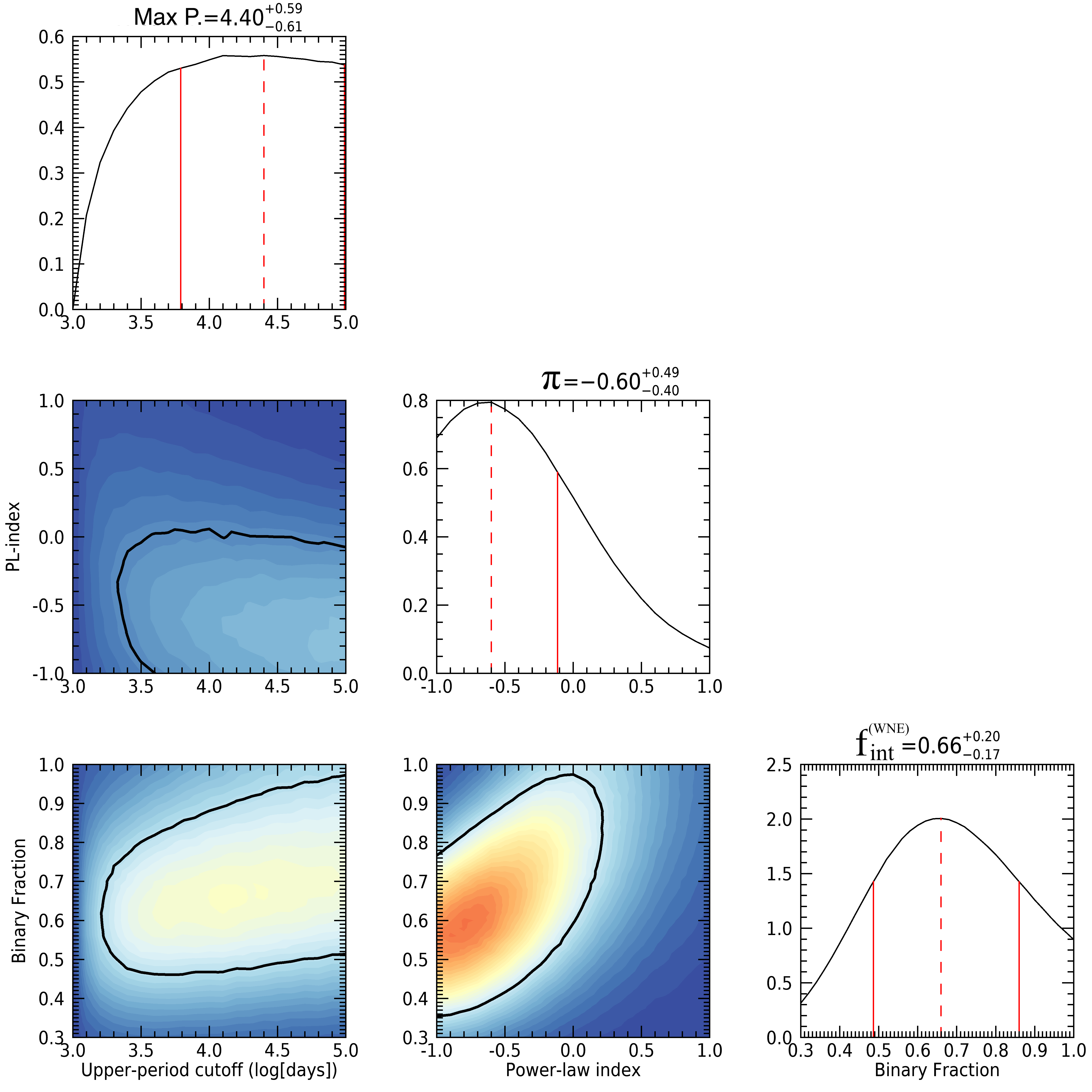}
    \caption{Posteriors for \logPmax{}, \fint{} , and $\pi$ for $\kappa = -1.0$. For each posterior, the solid red lines show HDI68, and the dashed red line shows the mode.}
    \label{fig:kappa_minus1}
\end{figure}

Here we present the three-dimensional likelihood plots for the Bayesian analysis of the WNE sample with the assumptions and setup as described in \ref{sect:bayesian} while exploring values for the mass-ratio index ($\kappa$) of $+1.0$ (Fig. \ref{fig:kappa_plus1}) and $-1.0$ (Fig. \ref{fig:kappa_minus1}). The conclusions do not vary significantly as compared to a flat mass ratio ($\kappa=0.0$, Fig. \ref{fig:flat_pdist_binfrac}). 

\subsection{WC sample}
As discussed in Sect. \ref{sect:WC_bayesian}, we performed MC simulations with a Bayesian framework for the 12 Galactic WC stars in \citetalias{2020Dsilva} over four parameters, \logPmin{}, \logPmax{}, \fint{} , and $\pi$ , with the same step size as for the WNE sample and over the ranges displayed in Fig.~\ref{fig:WC_posteriors}. We calculated the four-dimensional likelihood and corresponding posteriors, assuming flat priors. With this, we enable a consistent comparison of the corresponding WC and WNE populations from an evolutionary perspective.

As in \citetalias{2020Dsilva}, we divided the stars into four \DelRV{} bins: 5\,$\le$\,\DelRV{}\,$\le$\,30\,\kms{} (six targets), 30\,$\le$\,\DelRV{}\,$\le$\,250\,\kms{} (no observations), 250\,$<$\,\DelRV{}\,$\le$\,300\,\kms{} (one observation), and  \DelRV{}\,$\ge$\,300\,\kms{} (no observations). We also enforced that our simulations have binaries in the following period bins: $P > 20$\,d (three systems), $P > 2000$\,d (two systems). With this setup, we simulated 10\,000 sets of 12 WC stars for each bin in this four-dimensional parameter space, hence $\sim 7\time10^9$ populations.

\end{document}